\documentclass[sigconf]{aamas} 

\usepackage{balance} 
\usepackage{subfig}
\usepackage[titletoc]{appendix}
\usepackage{bm}
\usepackage{tabularx}                             
\usepackage{algorithm}
\usepackage{algorithmicx}
\usepackage[noend]{algpseudocode}
\usepackage{amsmath}
\usepackage{listings}
\usepackage{enumerate}
\usepackage{pdfpages}
\usepackage{mathtools}
\usepackage{dsfont}

\usepackage{bbm}
\usepackage{dsfont}

\newcolumntype{L}[1]{>{\raggedright\arraybackslash}m{#1}}
\newcolumntype{P}[1]{>{\centering\arraybackslash}p{#1}}

\setcopyright{none}
\renewcommand\footnotetextcopyrightpermission[1]{}
\acmConference[AAMAS '24]{Proc.\@ of the 23rd International Conference
on Autonomous Agents and Multiagent Systems (AAMAS 2024)}{May 6 -- 10, 2024}
{Auckland, New Zealand}{N.~Alechina, V.~Dignum, M.~Dastani, J.S.~Sichman (eds.)}
\copyrightyear{2024}
\acmYear{2024}
\acmDOI{}
\acmPrice{}
\acmISBN{}

\acmSubmissionID{953}

\title[Confidence-Based Curriculum Learning for Multi-Agent Path Finding]{Confidence-Based Curriculum Learning for\\Multi-Agent Path Finding}

\author{Thomy Phan}
\affiliation{
  \institution{University of Southern California}
  \city{Los Angeles}
  \country{USA}}
\email{thomy.phan@usc.edu}

\author{Joseph Driscoll}
\affiliation{
  \institution{Georgia Institute of Technology}
  \city{Atlanta}
  \country{USA}}
\email{jdriscoll7@gatech.edu}

\author{Justin Romberg}
\affiliation{
  \institution{Georgia Institute of Technology}
  \city{Atlanta}
  \country{USA}}
\email{jrom@ece.gatech.edu}

\author{Sven Koenig}
\affiliation{
  \institution{University of Southern California}
  \city{Los Angeles}
  \country{USA}}
\email{skoenig@usc.edu}

\begin{abstract}
A wide range of real-world applications can be formulated as \emph{Multi-Agent Path Finding (MAPF)} problem, where the goal is to find collision-free paths for multiple agents with individual start and goal locations. State-of-the-art MAPF solvers are mainly centralized and depend on global information, which limits their scalability and flexibility regarding changes or new maps that would require expensive replanning. \emph{Multi-agent reinforcement learning (MARL)} offers an alternative way by learning decentralized policies that can generalize over a variety of maps. While there exist some prior works that attempt to connect both areas, the proposed techniques are heavily engineered and very complex due to the integration of many mechanisms that limit generality and are expensive to use. We argue that much simpler and general approaches are needed to bring the areas of MARL and MAPF closer together with significantly lower costs.
In this paper, we propose \emph{Confidence-based Auto-Curriculum for Team Update Stability (CACTUS)} as a lightweight MARL approach to MAPF. CACTUS defines a simple reverse curriculum scheme, where the goal of each agent is randomly placed within an allocation radius around the agent's start location. The allocation radius increases gradually as all agents improve, which is assessed by a confidence-based measure.
We evaluate CACTUS in various maps of different sizes, obstacle densities, and numbers of agents. Our experiments demonstrate better performance and generalization capabilities than state-of-the-art MARL approaches with less than 600,000 trainable parameters, which is less than 5\% of the neural network size of current MARL approaches to MAPF.
\end{abstract}


\newcommand{\BibTeX}{\rm B\kern-.05em{\sc i\kern-.025em b}\kern-.08em\TeX}

\begin{document}


\pagestyle{fancy}
\fancyhead{}


\maketitle 


\section{Introduction}

\begin{figure}[!ht]
	\centering
	\includegraphics[width=0.45\textwidth]{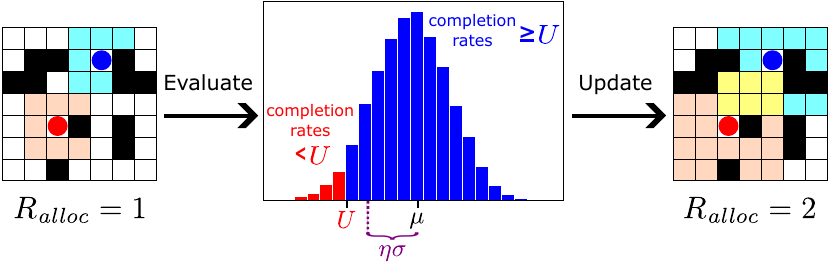}
     \caption{Curriculum update scheme of CACTUS. The agents (colored circles) are trained and evaluated w.r.t. a goal allocation radius $R_{\textit{alloc}}$ (shaded squares around the agents). When the average completion rate $\mu$ exceeds the decision threshold $U$ with a certain confidence level such that $\mu - \eta \sigma \geq U$, the allocation radius $R_{\textit{alloc}}$ is incremented by 1.}
     \Description{Curriculum update scheme of CACTUS. The agents (colored circles) are trained and evaluated w.r.t. a goal allocation radius $R_{\textit{alloc}}$ (shaded squares around the agents). When the average completion rate $\mu$ exceeds the decision threshold $U$ with a certain confidence level such that $\mu - \eta \sigma \geq U$, the allocation radius $R_{\textit{alloc}}$ is incremented by 1.}
     \label{fig:cactus_scheme}
\end{figure}

A wide range of real-world applications like goods transportation in warehouses,  search and rescue missions, and traffic management can be formulated as \emph{Multi-Agent Path Finding (MAPF)} problem, where the goal is to find collision-free paths for multiple agents with individual start and goal locations. Finding optimal solutions w.r.t. flowtime or makespan is NP-hard \cite{stern2019multi,ratner1986finding}. Despite the problem complexity, there exist a variety of MAPF solvers that find optimal \cite{sharon2012conflict}, bounded suboptimal \cite{cohen2016bounded}, or quick feasible solutions \cite{li2021anytime}. Most MAPF solvers are centralized and require global information, which limits scalability and flexibility regarding changes that would need expensive replanning. This also limits applicability to partially observable real-time domains \cite{sartoretti2019primal}.

\emph{Multi-agent reinforcement learning (MARL)} offers an alternative way by learning decentralized policies that can generalize over a variety of maps and make decisions under partial observability \cite{bucsoniu2010multi,tan1993multi}. State-of-the-art MARL algorithms are based on \emph{centralized training for decentralized execution (CTDE)}, where training takes place in a laboratory or a simulator with access to global information to learn coordinated policies that can be executed independently under partially observability afterwards \cite{lowe2017multi,rashid2018qmix}.

MAPF and MARL have been very active research areas in the last few years with impressive advances on both sides, resulting in a variety of sophisticated algorithms \cite{cohen2016bounded,li2021lifelong,vinyals2019grandmaster,jaderberg2019human}. Despite these advances, both fields have been mainly studied independently of each other. However, MARL could benefit MAPF in various ways:
\begin{enumerate}
\item \textbf{Efficiency}: The learned policies are reactive and decentralized therefore alleviating the computational and communication requirements of centralized MAPF solvers \cite{rashid2018qmix}.
\item \textbf{Generalization}: The learned policies can generalize over a variety of maps thus do not require complete retraining or replanning when being used on new maps \cite{sartoretti2019primal}.
\item \textbf{Robustness}: The learned policies make decisions based on actual observations therefore being able to react to local changes, i.e., emerging obstacles or new paths, without requiring replanning of the whole system \cite{phan2021resilient}.
\end{enumerate}

On the other hand, MAPF poses an exciting challenge for MARL due to its practical relevance and the following aspects \cite{sharon2012conflict,silver2005cooperative}:
\begin{enumerate}
\item \textbf{Sparse Rewards}: MAPF represents a complex navigation problem, where all agents are only rewarded for reaching their goals. Naive MARL would need exhaustive exploration to obtain informative data, which is time-consuming \cite{florensa2017reverse}.
\item \textbf{Dynamic Constraints}: Agents are not allowed to collide therefore having temporal constraints in addition to static constraints imposed by obstacles and boundaries \cite{silver2005cooperative}.
\item \textbf{Coordination}: MAPF requires coordination of spatially close agents with potentially emergent effects like congestion or circulation. So far, most MARL methods only focus on coordination on a small scale though \cite{lowe2017multi,yu2022surprising}. 
\end{enumerate}

We believe that addressing MAPF via MARL can provide a fruitful research direction that would benefit both areas. While there are prior works that attempt to connect these areas, the proposed techniques are heavily engineered and very complex, using extremely large neural networks, extensively shaped rewards, and centralized MAPF solvers for imitation learning \cite{sartoretti2019primal,damani2021primal,wang2023scrimp}. We argue that much simpler and general approaches are needed to bring the areas of MARL and MAPF closer together with significantly lower costs.

In this paper, we propose \emph{Confidence-based Auto-Curriculum for Team Update Stability (CACTUS)} as a lightweight MARL approach to MAPF. CACTUS defines a simple reverse curriculum scheme, where the goal of each agent is randomly placed within an allocation radius around the agent's start location. The allocation radius increases gradually as all agents improve, which is assessed by a confidence-based measure as shown in Fig. \ref{fig:cactus_scheme}. Our contributions are as follows:
\begin{itemize}
\item We formulate the MAPF problem as a straightforward stochastic game with automatic collision prevention and sparse rewards to solve it in a black-box manner, which is more general and intuitive for standard MARL methods.
\item Based on the stochastic game formulation, we propose a simple reverse curriculum scheme that gradually increases the potential distance between start and goal locations to enhance state-of-the-art MARL techniques that would likely fail to learn any meaningful policy otherwise.
\item We evaluate CACTUS in various maps of different sizes, obstacle densities, and numbers of agents. Our experiments demonstrate better performance and generalization capabilities than state-of-the-art MARL approaches with less than 600,000 trainable parameters, which is less than 5\% of the neural network size of current MARL approaches to MAPF.
\end{itemize}

\section{Background}

\subsection{Multi-Agent Path Finding}\label{subsec:mapf_background}

We focus on \emph{maps} as undirected unweighted \emph{graphs} $G = \langle \mathcal{V}, \mathcal{E} \rangle$, where vertex set $\mathcal{V}$ contains all possible locations and edge set $\mathcal{E}$ contains all possible transitions or movements between adjacent locations. An \emph{instance} $I$ consists of a map $G$ and a set of \emph{agents} $\mathcal{D} = \{1, ..., N\}$ with each agent $i \in \mathcal{D}$ having a \emph{start location} $v_{\textit{start},i} \in \mathcal{V}$ and a \emph{goal location} $v_{\textit{goal},i} \in \mathcal{V}$. We assume that $v_{\textit{start},i} \neq v_{\textit{start},j}$ and $v_{\textit{goal},i} \neq v_{\textit{goal},j}$ for any agent pair $i,j \in \mathcal{D}$ with $i \neq j$.

MAPF aims to find collision-free plans for all agents. A \emph{plan} $P = \{ p_1, ..., p_N \}$ consists of individual paths $p_i = \langle p_{i,0}, ..., p_{i,l(p_i)} \rangle$ per agent $i \in \mathcal{D}$, where $\langle p_{i,t}, p_{i,t+1} \rangle = \langle p_{i,t+1}, p_{i,t} \rangle \in \mathcal{E}$, $p_{i,0} = v_{\textit{start},i}$, $p_{i,l(p_i)} = v_{\textit{goal},i}$, and $l(p_i)$ is the \emph{length} or \emph{travel distance} of path $p_i$.

We consider \emph{vertex conlicts} $\langle a_i, a_j, v, t \rangle$ that occur when two agents $i, j \in \mathcal{D}$ occupy the same location $v \in \mathcal{V}$ at time step $t$ and \emph{edge conflicts} $\langle i, j, u, v, t \rangle$ that occur when two agents $i, j \in \mathcal{D}$ traverse the same edge $\langle u, v \rangle = \langle v, u \rangle \in \mathcal{E}$ in opposite directions at time step $t$ \cite{stern2019multi}. A plan $P$ is a \emph{solution}, i.e., \emph{feasible}, when it does not have any vertex or edge conflict therefore no collisions. Our goal is to find a solution $P^{*}$ that minimizes the \emph{flowtime} $\sum_{p \in P} l(p)$.

Despite MAPF being an NP-hard problem, there exist a variety of MAPF solvers that find optimal \cite{sharon2012conflict}, bounded suboptimal \cite{cohen2016bounded}, or quick feasible solutions \cite{li2021anytime}. Most MAPF solvers are centralized and require global information which limits scalability and flexibility regarding changes or new maps that would need expensive replanning and redistribution of plans. 

\subsection{\textbf{Multi-Agent Reinforcement Learning}}\label{subsec:marl_background}
MARL problems can be formulated as a partially observable \emph{stochastic game} $\mathcal{M} = \langle \mathcal{D},\mathcal{S},\mathcal{A},\mathcal{P},\mathcal{R},\mathcal{Z},\Omega \rangle$, where $\mathcal{D} = \{1,...,N\}$ is a set of agents, $\mathcal{S}$ is a set of states $s_{t}$, $\mathcal{A} = \mathcal{A}_{1} \times ... \times \mathcal{A}_{N}$ is the set of joint actions $a_{t} = \langle a_{t,1},...,a_{t,N} \rangle$, $\mathcal{P}(s_{t+1}|s_{t}, a_{t})$ is the transition probability, $\mathcal{R}(s_{t},a_{t}) = \langle r_{t,1},...,r_{t,N} \rangle \in \mathbb{R}^{N}$ is the joint reward with $r_{t,i}$ being the reward of agent $i \in \mathcal{D}$, $\mathcal{Z}$ is a set of local observations $z_{t,i}$ for each agent $i$, and $\Omega(s_{t+1}) = z_{t+1} = \langle z_{t+1,1},...,z_{t+1,N} \rangle \in \mathcal{Z}^{N}$ is the subsequent joint observation. Each agent $i$ maintains an action-observation \emph{history} $\tau_{t,i} \in (\mathcal{Z} \times \mathcal{A}_{i})^{t}$. $\pi = \langle\pi_{1},...,\pi_{N}\rangle$ is the \emph{joint policy} with \emph{local policies} $\pi_i$, where $\pi_{i}(a_{t,i}|\tau_{t,i})$ is the action selection probability of agent $i$.
Local policy $\pi_{i}$ can be evaluated with a \emph{value function} $Q_{i}^{\pi}(s_{t},a_{t}) = \mathbb{E}_{\pi}[R_{t,i}|s_{t},a_{t}]$ for all states $s_{t} \in \mathcal{S}$ and $a_{t} \in \mathcal{A}$, where $R_{t,i} = \sum_{k=0}^{T-1} \gamma^{k} r_{t+k,i}$ is the \emph{return} of agent $i$, $T > 0$ is the \emph{horizon}, and $\gamma \in [0,1]$ is the \emph{discount factor}. 

In cooperative MARL, the goal is to find an \emph{optimal joint policy} $\pi^{*} = \langle \pi^{*}_1, ..., \pi^{*}_N \rangle$ that maximizes the \emph{utilitarian metric} for all states $s_{t} \in \mathcal{S}$:
\begin{equation}\label{eq:utilitarian_metric}
Q_{\textit{tot}}^{\pi}(s_{t}, a_{t}) = \sum_{i \in \mathcal{D}}Q^{\pi}_{i}(s_{t}, a_{t})
\end{equation}

\subsubsection{\textbf{Policy Gradient MARL}}

To learn optimal policies $\pi^{*}_i$ in large state spaces, function approximators $\hat{\pi}_{i,\theta}$ with parameters $\theta$ are trained with gradient ascent on an estimate of $J = \mathbb{E}_{\pi}[R_{0,i}]$. \emph{Policy gradient methods} use gradients $g$ of the following form \cite{sutton2000policy}:
\begin{equation}\label{eq:policy_gradients}
g = A_{i}^{\hat{\pi}}(s_{t},a_{t})\nabla_{\theta} \textit{log} \hat{\pi}_{i,\theta}(a_{t,i}|\tau_{t,i})
\end{equation}
where $A_{i}^{\hat{\pi}}(s_{t},a_{t}) = Q_{i}^{\hat{\pi}}(s_{t},a_{t}) - V_{i}^{\hat{\pi}}(s_{t})$ is the \emph{advantage} of agent $i$ and $V_{i}^{\hat{\pi}}(s_{t}) = \mathbb{E}_{\pi}[R_{t,i}|s_{t}]$ is its state value function. \emph{Actor-critic} approaches often approximate $\hat{A_{i}} \approx A_{i}^{\hat{\pi}_{i}}$ by replacing $Q_{i}^{\hat{\pi}}(s_{t},a_{t})$ with $R_{t,i}$ and $V_{i}^{\hat{\pi}}$ with $\mathbb{E}_{\pi_{i}}[Q_{i}^{\hat{\pi}}]$. $Q_{i}^{\hat{\pi}}$ can be approximated with a \emph{critic} $\hat{Q}_{i,\omega}$ and parameters $\omega$ using value-based RL \cite{watkins1992q, mnih2015human}.

Alternatively, $\hat{\pi}_{i,\theta}$ can be trained via \emph{proximal policy optimization (PPO)} by iteratively minimizing the following surrogate loss \cite{schulman2017proximal}:
\begin{equation}\label{eq:ppo_loss}
\mathcal{L}^{\textit{PPO}}_{i}(\theta) = \mathbb{E}[\textit{min}\{\hat{A_{i}}\phi_{t,i}(\theta), \hat{A_{i}}\textit{clip}(\phi_{t,i}(\theta), 1 - \epsilon, 1 + \epsilon)\}]
\end{equation}
where $\phi_{t,i}(\theta) = \frac{\hat{\pi}_{i,\theta}(a_{t,i}|\tau_{t,i})}{\hat{\pi}^{\textit{old}}_{i,\theta}(a_{t,i}|\tau_{t,i})}$ is the policy probability ratio and $\epsilon \in [0,1)$ is a clipping parameter. For simplicity, we omit the parameters $\theta$, $\omega$ and write $\hat{\pi}_{i}$, $\hat{Q}_{i}$ for the rest of the paper.

\subsubsection{\textbf{Centralized Training Decentralized Execution (CTDE)}}\label{subsec:ctde}

For many problems, training takes place in a laboratory or in a simulated environment, where global information is available \cite{rashid2018qmix}. Therefore, state-of-the-art MARL algorithms approximate value functions $\hat{Q}_{i}$, which condition on global states $s_{t}$ and joint actions $a_{t}$, and use them as critic in Eq. \ref{eq:policy_gradients} or \ref{eq:ppo_loss} \cite{lowe2017multi,yu2022surprising}. While the value functions $\hat{Q}_{i}$ are only required during training, the learned policies $\hat{\pi}_{i}$ only condition on local histories $\tau_{t,i}$ thus being independently executable. Unlike MAPF, these policies can \emph{generalize} over a variety of scenarios and thus ideally do not need any centralized retraining or replanning for changes or new maps \cite{sartoretti2019primal}.

$\hat{Q}_{i}$ can be approximated separately for each agent $i$ while integrating global information, which is done in actor-critic MARL algorithms like MAPPO or MADDPG \cite{lowe2017multi,yu2022surprising}. However, this approach lacks a \emph{multi-agent credit assignment mechanism} for agent teams, where all agents optimize the same objective $Q_{\textit{tot}}$ (Eq. \ref{eq:utilitarian_metric}).

Alternatively, a common value function $\hat{Q}(\tau_{t}, a_{t}) \approx Q_{\textit{tot}}(s_{t}, a_{t})$ can be learned, which is factorized into $\langle \hat{Q}_{1}, ..., \hat{Q}_{N} \rangle$ as \emph{local utility functions} by using a \emph{factorization operator} $\Psi$ \cite{phan2021vast,phan2023attention}:
\begin{equation}\label{eq:vff}
\hat{Q}(\tau_{t}, a_{t}) = \Psi(\hat{Q}_{1}(\tau_{t,1},a_{t,1}),...,\hat{Q}_{N}(\tau_{t,N},a_{t,N}))
\end{equation}
In practice, $\Psi$ is realized with deep neural networks, such that $\langle \hat{Q}_{1}, ..., \hat{Q}_{N} \rangle$ can be learned end-to-end via backpropagation by minimizing the mean squared \emph{temporal difference (TD)} error \cite{rashid2018qmix,sunehag2017value}. A factorization operator $\Psi$ is \emph{decentralizable} when satisfying the \emph{IGM (Individual-Global-Max)} such that \cite{son2019qtran}:

\begin{equation}\label{eq:igm}
\textit{argmax}_{\mathbf{a_{t}}}\hat{Q}(\tau_{t}, a_{t}) =
\begin{pmatrix} 
\textit{argmax}_{a_{t,1}}\hat{Q}_{1}(\tau_{t,1},a_{t,1})\\
\vdots \\
\textit{argmax}_{a_{t,N}}\hat{Q}_{N}(\tau_{t,N},a_{t,N})
\end{pmatrix}
\end{equation}

There exists a variety of factorization operators $\Psi$, which satisfy Eq. \ref{eq:igm} using monotonicity constraints like QMIX \cite{rashid2018qmix} or nonlinear transformation like QPLEX or QTRAN \cite{son2019qtran,wang2020qplex}.

\subsection{Curriculum Learning}

\emph{Curriculum learning} is a machine learning paradigm, inspired by human learning, to master complex tasks through stepwise solving of easier (sub-)tasks, which are sorted by difficulty \cite{bengio2009curriculum,soviany2022curriculum}. The difficulty can depend on various aspects like the complexity of data samples, the objective function, or the learned model \cite{narvekar2016source,florensa2017reverse}.
Curriculum learning has been applied to \emph{reinforcement learning (RL)} to solve hard exploration problems with sparse rewards or dynamic constraints \cite{narvekar2020curriculum}. The methods are typically based on self-play \cite{silver2017mastering,tesauro1995temporal}, task graphs with traversal mechanisms \cite{silva2018object}, or automatic generation of tasks \cite{gabor2019scenario,dennis2020emergent}.

A key challenge of curriculum learning is to find or generate a suitable sequence of tasks that are neither too easy nor too difficult for the learner to ensure steady and robust progress \cite{florensa2017reverse,silva2018object,gabor2019scenario}.

We focus on \emph{reverse curriculum} learning, where we assume explicit goal states as in the MAPF problem (Section \ref{subsec:mapf_background}). The curriculum consists of a sequence of tasks, where the (expected) distance between agent and goal gradually increases \cite{asada1996purposive,florensa2017reverse}.

\section{Related Work}

\paragraph{\textbf{Reverse Curriculum Generation}}
Many works on RL-based motion control assume a single goal state, which is easy to specify \cite{mcaleer2018solving,agostinelli2019solving}. A \emph{reverse curriculum} is defined, where the start state is initialized within a short distance to the goal state. The distance gradually increases with the convergence or performance improvement of the agent \cite{asada1996purposive,florensa2017reverse}. Our work is based on reverse curriculum generation, focusing on \emph{multi-agent path finding (MAPF)}. In MAPF, there are \emph{several goal states} that are unique per instance $I$ (which can vary for the same map $G$ though). We propose a simple \emph{confidence-based approach} to adapt the curriculum by considering the \emph{uncertainty} of performance estimates.

\paragraph{\textbf{Curriculum Learning in MARL}}

Curriculum learning has been widely used in single-agent or two-player zero-sum games to improve convergence speed or performance. While many of these approaches are based on foundations of multi-agent learning \cite{sukhbaatar2018intrinsic,dennis2020emergent,du2021takes}, there exist methods particularly designed for MARL based on self-play, agent skills, and population-based training \cite{jaderberg2019human,vinyals2019grandmaster,wu2023portal}. These methods are typically very complex due to heavily engineered architectures and mechanisms thus requiring a significant amount of compute. As our work focuses on \emph{simple and efficient} MARL approaches to MAPF, we do not consider such resource and tuning-intensive training regimes.

\paragraph{\textbf{MARL for MAPF}}

MAPF and MARL are very active research areas with remarkable progress in recent years \cite{li2021anytime,vinyals2019grandmaster,jaderberg2019human}. Both fields have been mainly studied independently of each other, with only a few works attempting to connect them. The first work in this direction is \emph{PRIMAL}. PRIMAL and its successor approaches are heavily engineered and very complex, using extremely large neural networks, extensively shaped rewards, and centralized MAPF solvers for imitation learning to address the challenging aspects of MAPF, i.e., sparse rewards, dynamic constraints, and coordination \cite{sartoretti2019primal,damani2021primal,wang2023scrimp}. Despite their effectiveness, these approaches are very expensive to use due to significant effort on fine-tuning and enormous computational and data requirements. Some recent works proposed manually designed curricula to enhance PRIMAL but still rely on very complex architectures and reward functions \cite{zhao2023curriculum,pham2023crowd}.
Besides applying MARL to MAPF, there have been other attempts to combine MAPF with machine learning techniques to guide or select centralized search algorithms \cite{HuangAAAI22,kaduri2020algorithm,huang2021learning,phanAAAI24}.
Our goal is to provide a suitable foundation to bring the areas of MARL and MAPF closer together with \emph{significantly lower costs}. Therefore, we propose a \emph{simple reverse curriculum scheme} to ease applicability and enable faster progress in this direction.

\section{MAPF as a Stochastic Game}\label{sec:mapf_stochastic_game}

To apply MARL techniques to MAPF in a general way, we first need to formulate the MAPF problem defined in Section \ref{subsec:mapf_background} as a stochastic game $\mathcal{M}$ according to Section \ref{subsec:marl_background}. Similar to prior work \cite{sartoretti2019primal,damani2021primal,wang2023scrimp,stern2019multi}, we focus on discrete gridworlds but try to keep our formulation general. An adaptation of our methods to arbitrary graphs, e.g., using graph neural networks, is left for future work.

\begin{figure}[!ht]
	\centering
	\includegraphics[width=0.35\textwidth]{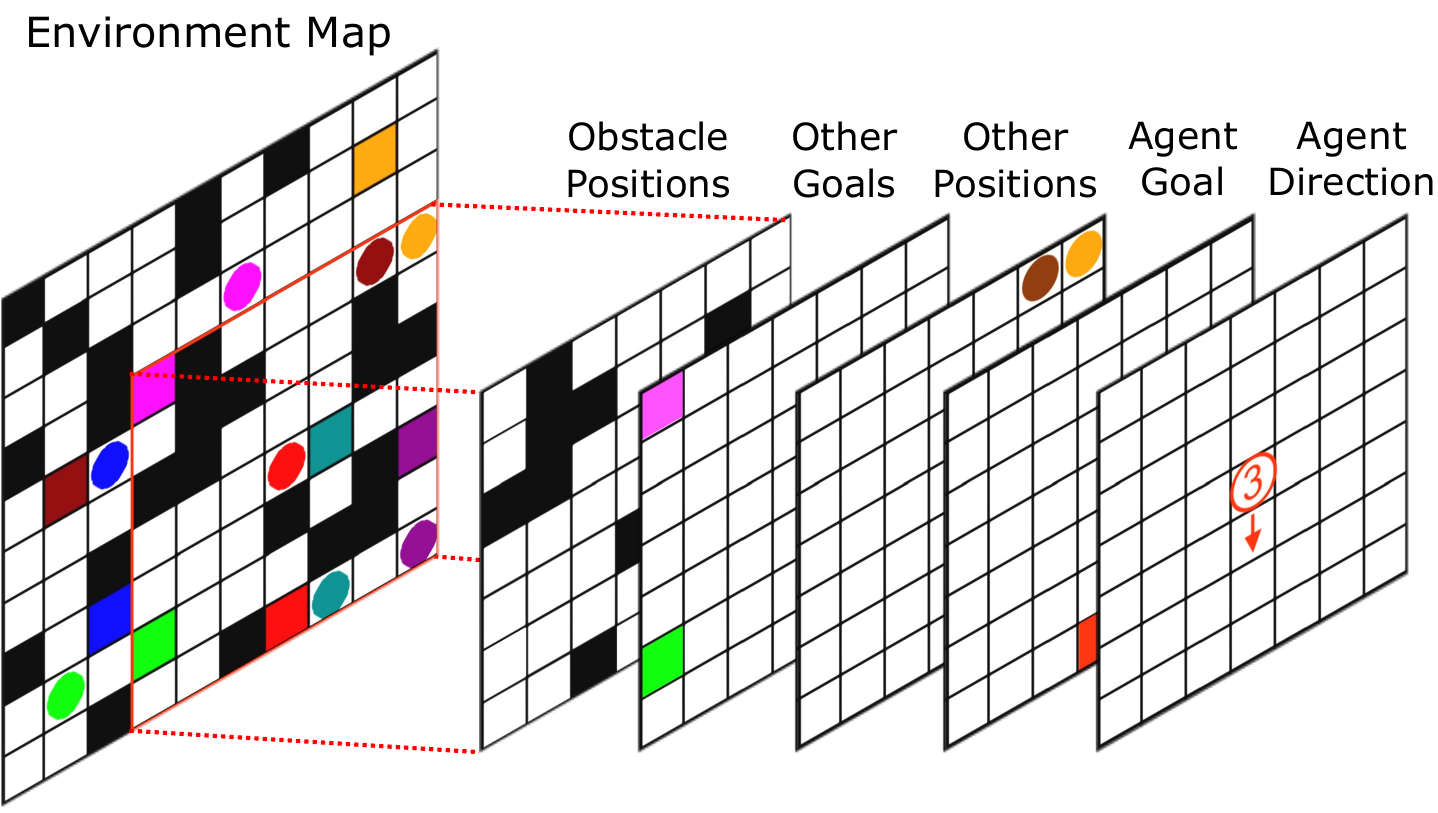}
     \caption{Example for an individual observation of the red agent in a gridworld domain. Agents are represented as colored circles, their goals as similarly-colored squares, and obstacles as black squares. Each agent $i$ has a limited \emph{field of view (FOV)} of the environment map, which is centered around its location encoded by five channels: locations of obstacles, location of other agents' goals, locations of nearby agents, and location of the goal $v_{\textit{goal},i}$ if within the FOV, and the Manhattan distance and direction of agent $i$ to its goal.}
     \Description{Example for an individual observation of the red agent in a gridworld domain. Agents are represented as colored circles, their goals as similarly-colored squares, and obstacles as black squares. Each agent $i$ has a limited \emph{field of view (FOV)} of the environment map, which is centered around its location encoded by five channels: locations of obstacles, location of other agents' goals, locations of nearby agents, and location of the goal $v_{\textit{goal},i}$ if within the FOV, and the Manhattan distance and direction of agent $i$ to its goal.}
     \label{fig:observation_example}
\end{figure}

In both settings, the set of agents $\mathcal{D}$ is equivalent.
Given a map $G = \langle \mathcal{V}, \mathcal{E} \rangle$, the state space $\mathcal{S}$ is defined by the joint locations of all agents $s_t = \langle v_{t,i}, ..., v_{t,N} \rangle \in \mathcal{S} \subset \mathcal{V}^N$, where each location in $s_t$ is unique such that $v_{t,i} \neq v_{t,j}$ for each agent pair $i,j \in \mathcal{D}$ with $i \neq j$. The individual action space $\mathcal{A}_{i}$ of each agent $i$ is defined by the maximum degree of map $G$ plus a wait action. In 4-neighborhood gridworlds as displayed in Fig. \ref{fig:observation_example}, each agent would be able to wait or move in all cardinal directions. The state transitions are deterministic, where a valid move action will change the location of the corresponding agent. Attempts to move over non-existent edges or cause vertex or edge conflicts, i.e., collisions, are automatically treated as wait action. The individual reward $r_{t,i}$ is defined by +1 if agent $i$ reaches its goal $v_{\textit{goal},i}$, 0 when agent $i$ is staying at its goal location $v_{\textit{goal},i}$, and -1 otherwise. Each agent $i$ can partially observe the state $s_t$ through a local neighborhood around its location $v_{t,i}$. For gridworlds, we assume a $7 \times 7$ \emph{field of view (FOV)} similar to PRIMAL, which is illustrated in Fig. \ref{fig:observation_example} \cite{sartoretti2019primal}. The features of an observation $z_{t,i}$, i.e., obstacles, other agents and their goals, and the direction and goal location of agent $i$, are encoded as a multi-channel image. The direction channel encodes the Manhattan distance to the goal $v_{\textit{goal},i}$ and indicates the direction to it.

When the discount factor is $\gamma = 1$, the negated return $-R_{t,i}$ of each agent $i$ is equivalent to its travel distance $l(p_i)$ from time step $t$, if $v_{\textit{goal},i}$ was reached, and horizon $T$ otherwise. Therefore, maximizing $Q_{\textit{tot}} = \sum_{i \in \mathcal{D}}Q_{i} = -\mathbb{E}_{I}[\sum_{p \in P} l(p)]$ in MARL (Section \ref{subsec:marl_background} and Eq. \ref{eq:utilitarian_metric}) would be equivalent to minimizing the expected flowtime in MAPF w.r.t. any instance $I$ on map $G$ (Section \ref{subsec:mapf_background}).

Since any time step is penalized with -1 anyway (unless an agent reaches or occupies its goal), all agents are discouraged from unnecessary delays, which includes collision attempts.
Unlike prior work, we do not need additional penalties for particular situations like collisions, blocking, or waiting which could fundamentally change the actual objective and lead to unintended side-effects \cite{skalse2022defining}

Thus, our problem formulation is simpler and more general, which allows us to solve it in a black-box manner that is more intuitive for standard MARL methods \cite{rashid2018qmix,wang2020qplex,yu2022surprising}. However, the simplicity of our formulation notably increases difficulty since the reward is sparse in contrast to PRIMAL and related approaches.

\section{Confidence-based Curriculum}

\subsection{Training Scheme}\label{subsec:training_scheme}

\begin{figure}[!ht]
	\centering
	\includegraphics[width=0.45\textwidth]{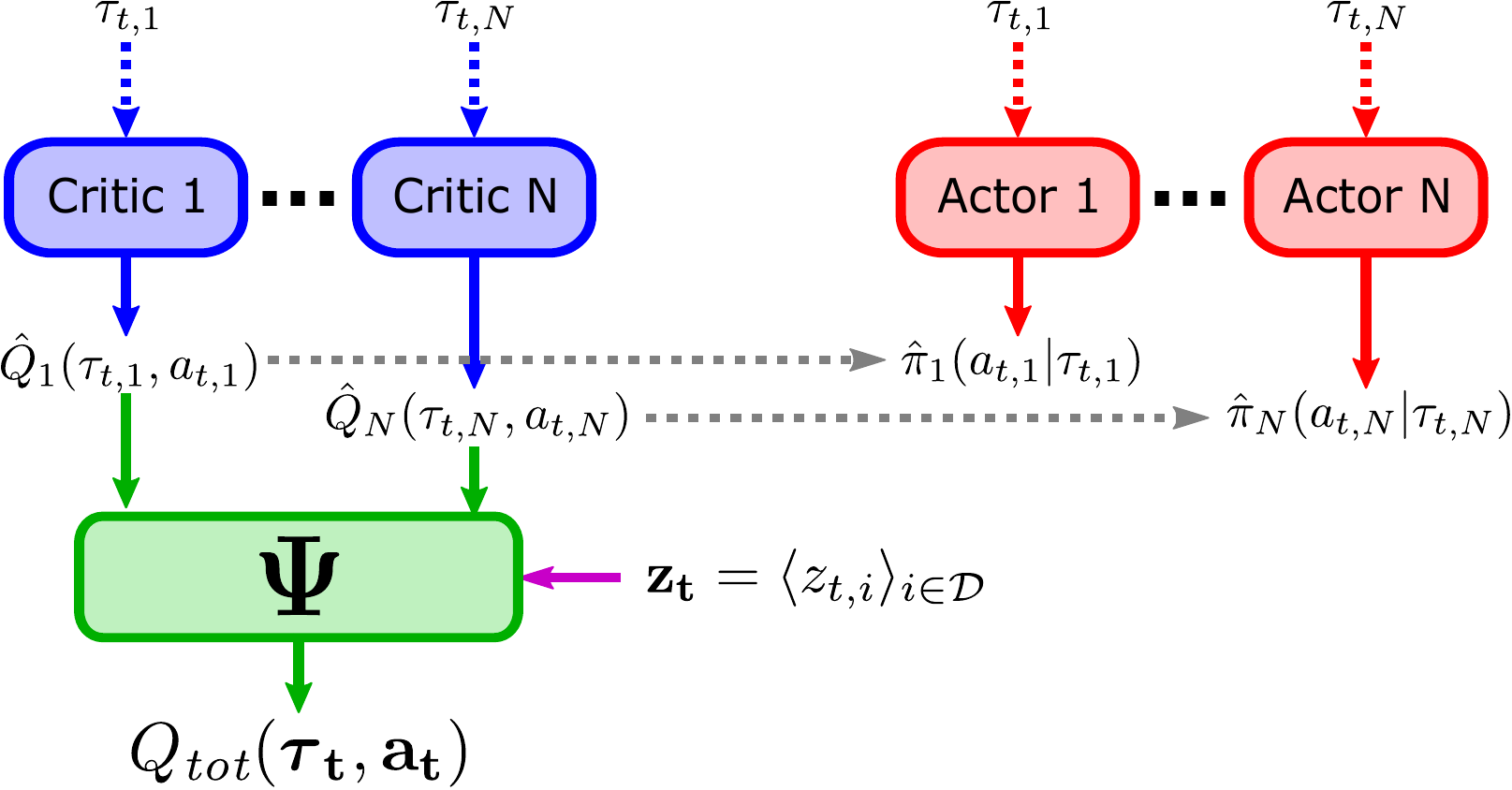}
     \caption{Common actor-critic scheme as used in various prior work on cooperative MARL \cite{su2021value,phan2021resilient}. A separate critic is trained for each actor using some centralized factorization operator $\Psi$ like QMIX or QPLEX \cite{phan2021vast}.}
     \Description{Common actor-critic scheme as used in various prior work on cooperative MARL \cite{su2021value,phan2021resilient}. A separate critic is trained for each actor using some centralized factorization operator $\Psi$ like QMIX or QPLEX \cite{phan2021vast}.}
     \label{fig:training_scheme}
\end{figure}

We assume a separate training phase to learn coordinated local policies $\hat{\pi}_{i}$ for decentralized execution. We train $\hat{\pi}_{i}$ via policy gradient methods according to Eq. \ref{eq:policy_gradients} or \ref{eq:ppo_loss}. The critics $\hat{Q}_{i}$ are trained via CTDE methods to exploit global information during training using either independent learning like MAPPO or value factorization like QMIX or QPLEX as illustrated in Fig. \ref{fig:training_scheme} \cite{yu2022surprising,rashid2018qmix,wang2020qplex}. Since the value factorization based actor-critic scheme has been used in a variety of prior work \cite{su2021value,phan2021resilient}, we do not claim novelty here, but propose it as a basic approach to train cooperative policies via credit assignment mechanisms \cite{rashid2018qmix,wang2020qplex,son2019qtran}.

To address the coordination problem as mentioned in the introduction, we suggest to optimize individual utilities $\hat{Q}_{i} \approx Q_{i}^{\hat{\pi}}$ under consideration of the utilitarian metric $Q_{\textit{tot}}$ in Eq. \ref{eq:utilitarian_metric}. For that, the individual utilities $\hat{Q}_{i}$ can be learned end-to-end through a factorization operator $\Psi$ like QMIX or QPLEX to consider multi-agent credit assignment from a cooperative perspective \cite{rashid2018qmix,wang2020qplex,phan2021vast,phan2023attention}. The factorization operator $\Psi$ approximates the expected sum of individual returns by minimizing the \emph{factorization loss} $\mathcal{L}^{\Psi}$ defined by:
\begin{equation}\label{eq:factorization_loss}
\mathcal{L}^{\Psi} = \mathbb{E}\Big[\Big(\Psi(\hat{Q}_{1}(\tau_{t,1},a_{t,1}),...,\hat{Q}_{N}(\tau_{t,N},a_{t,N})) - \sum_{i \in \mathcal{D}}R_{t,i}\Big)^2 \Big]
\end{equation}

The local policies $\hat{\pi}_{i}$ are then trained according to Eq. \ref{eq:policy_gradients} or \ref{eq:ppo_loss} using \emph{counterfactual advantages} $\hat{A_{i}}$ defined by \cite{su2021value}:
\begin{equation}
\hat{A_{i}}(\tau_{t,i},a_{t,i}) = R_{t,i} - \sum_{a' \in \mathcal{A}_{i}}\hat{\pi}_{i}(a'|\tau_{t,i})\hat{Q}_{i}(\tau_{t,i},a')
\end{equation}
where the return $R_{t,i}$ represents the negative travel distance $-l(p_i)$ of agent $i$ from time step $t$ according to Section \ref{sec:mapf_stochastic_game}. The advantage $\hat{A_{i}}$ incentivizes the optimization of travel distances under implicit consideration of other agents through $\hat{Q}_{i}$ and $\Psi$ w.r.t. Eq. \ref{eq:utilitarian_metric} and \ref{eq:factorization_loss}.

\subsection{Reverse Curriculum Scheme}

The training scheme described above represents a general approach to learn coordinated policies $\hat{\pi}_{i}$ \cite{su2021value,phan2021resilient}. However, sparse rewards and dynamic constraints in our MAPF setting (Section \ref{sec:mapf_stochastic_game}) pose particular challenges that require a suitable curriculum to learn meaningful policies \cite{asada1996purposive,florensa2017reverse}. Unlike prior work that relied on complex reward functions with various penalties and expensive expert data for imitation learning, we propose \emph{Confidence-based Auto-Curriculum for Team Update Stability (CACTUS)} to enhance the training scheme of Section \ref{subsec:training_scheme} without significant costs.

At the beginning of every episode $m$, each agent $i$ starts at a random location $v_{\textit{start},i} \in \mathcal{V}$ and needs to navigate to an assigned goal location $v_{\textit{goal},i} \in \mathcal{V}$ which is randomly placed within an \emph{allocation radius} $R_{\textit{alloc}}$\footnote{A vertex distance measure is required, which can depend on the number of edges, for example. In this paper, we measure the distance between two positions $(x_1, y_1)$ and $(x_2, y_2)$ by $\textit{max}\{|x_1 - x_2|, |y_1 - y_2|\}$ for two-dimensional environments. } around the start location $v_{\textit{start},i}$. $R_{\textit{alloc}}$ characterizes the \emph{potential difficulty} of generated instances $I$ as larger allocation radii may require the agents to move and explore over longer distances to locate their respective goals. Thus, our reverse curriculum scheme starts with a small allocation radius of $R_{\textit{alloc}} = 1$ and gradually increments $R_{\textit{alloc}}$ with improving performance, which is measured by the \emph{completion rate} $\rho^{\textit{complete}}_{m} = \frac{N^{\textit{goal}}_{m}}{N}$, where $N^{\textit{goal}}_{m} = |\{i \in \mathcal{D} | v_{t,i} = v_{\textit{goal},i}\}|$ is the number of agents that successfully reached their respective goals in episode $m$.

CACTUS uses a statistical approach to decide whether to increment $R_{\textit{alloc}}$ or not. After each epoch of $E$ episodes $m$, we measure the \emph{average completion rate} $\mu = \frac{1}{E} \sum^{E}_{m=1} \rho^{\textit{complete}}_{m}$ and its \emph{standard deviation} $\sigma = \sqrt{\frac{1}{E-1} \sum^{E}_{m=1} (\rho^{\textit{complete}}_{m} - \mu)^2}$.

Assuming that the completion rates $\rho^{\textit{complete}}_{m}$ follow a normal distribution, CACTUS increments $R_{\textit{alloc}}$ by 1, if $\mu - \eta \sigma \geq U$,
where $U \in (0,1)$ is the \emph{curriculum decision threshold} and $\eta > 0$ is a \emph{deviation factor} to specify the confidence level. For example, if $U = 75\%$ and $\eta = 2$ then $R_{\textit{alloc}}$ would be incremented only if all agents achieve an average completion rate over 75\% with a confidence level of around 97\%. Note that we only regard \emph{one-tailed tests} here, where we assume no upper limit to the average completion rate of agents (except for $\mu = 100\%$, where $\sigma$ would be zero). The curriculum update scheme is illustrated in Fig. \ref{fig:cactus_scheme}.

The complete formulation of CACTUS is given in Algorithm \ref{algorithm:CACTUS}. $\mathcal{G}$ is a set of training maps or a map generator, $\textit{DIST} : \mathcal{V} \times \mathcal{V} \rightarrow \mathbb{R}$ is a vertex distance function, $U$ is the curriculum decision threshold, and $\eta$ is the deviation factor.

\begin{algorithm}
\caption{Confidence-Based Curriculum Learning for MAPF}\label{algorithm:CACTUS}
\begin{algorithmic}[1]
\Procedure{$\textit{DRIVE}(\mathcal{G}, \textit{DIST}, U, \eta$)}{}
\State Initialize parameters of $\hat{\pi}_{i}, \hat{Q}_{i}$ for each agent $i \in \mathcal{D}$ and $\Psi$
\State Set $R_{\textit{alloc}} = 1$
\For{epoch $x \leftarrow 1,X$}
\For{episode $m \leftarrow 1,E$}
\State Randomly select or generate map $G$ from $\mathcal{G}$
\State Sample $s_{0}$ \Comment{Set start locations $v_{\textit{start},i}$}
\For{agent $i \in \mathcal{D}$}
\State Set $\tau_{0,i}$ based on $\Omega(s_{0})$
\State Set $\mathcal{V}_{\textit{goal,i}} \leftarrow \{v \in \mathcal{V} | \textit{DIST}(v, v_{\textit{start},i}) \leq R_{\textit{alloc}} \}$
\State Randomly select goal location $v_{\textit{goal},i}$ from $\mathcal{V}_{\textit{goal,i}}$ \Comment{Goal locations must be unique, i.e, $v_{\textit{goal},i} \neq v_{\textit{goal},j}$ if $i \neq j$}
\EndFor
\For{time step $t \leftarrow 0,T-1$}
\For{agent $i \in \mathcal{D}$}
	\State $a_{t,i} \sim \pi_{i}(\cdot|\tau_{t,i})$
\EndFor
\State $a_{t} \leftarrow \langle a_{t,1}, ..., a_{t,N} \rangle$
\State Execute joint action $a_{t}$
\State $s_{t+1} \sim \mathcal{T}(\cdot|s_{t}, a_{t})$ 
\State $z_{t+1} \leftarrow \Omega(s_{t+1})$
\State $e_{t} \leftarrow \langle \tau_{t}, a_{t}, r_{t}, z_{t+1}, \rangle$
\State Store experience sample $e_{t}$
\State $\tau_{t+1} \leftarrow \langle \tau_{t}, a_{t}, z_{t+1} \rangle$
\EndFor
\State $\rho^{\textit{complete}}_{m} \leftarrow |\{i \in \mathcal{D} | v_{t,i} = v_{\textit{goal},i}\}|/N$
\EndFor
\State Train $\Psi$ and $\hat{\pi}_{i}, \hat{Q}_{i}$ for each agent $i \in \mathcal{D}$ with all $e_{t}$
\State Calculate $\mu$ and $\sigma$ with all $\rho^{\textit{complete}}_{m}$
\If{$\mu - \eta \sigma \geq U$}\Comment{Curriculum update decision}
\State $R_{\textit{alloc}} \leftarrow R_{\textit{alloc}} + 1$
\EndIf
\EndFor
\State \Return $\langle \hat{\pi}_{1}, ..., \hat{\pi}_{N} \rangle$ 
\EndProcedure
\end{algorithmic}
\end{algorithm}

\subsection{Conceptual Discussion}

CACTUS represents a simple reverse curriculum scheme inspired by prior work \cite{asada1996purposive,florensa2017reverse}. Our work focuses on the MAPF problem, where we have multiple agents with different start and goal locations that can vary per instance $I$. To ensure generalization over a variety of MAPF instances and maps, our scheme adjusts the random allocation of goals around the random start locations.

In contrast to prior work \cite{zhao2023curriculum,pham2023crowd}, CACTUS does not separate learning of different skills like navigation and collision avoidance. As illustrated in Fig. \ref{fig:cactus_scheme}, all agents first need to focus on reaching their respective goals, which are allocated in close proximity within $R_{\textit{alloc}}$. With increasing $R_{\textit{alloc}}$, the allocation areas of different agents may overlap, automatically causing agents to interact with each other thus increasing coordination pressure. $R_{\textit{alloc}}$ is only incremented when agents are able to coordinate and reach their goals with sufficiently high confidence, which is checked with the hyperparameters $U$ and $\eta$. Thus, CACTUS offers an adaptive approach to solving MAPF problems via MARL without requiring explicit separation of agent skills \cite{zhao2023curriculum,pham2023crowd,vinyals2019grandmaster}, extensive engineering of rewards, or expensive acquisition of expert data \cite{sartoretti2019primal,damani2021primal,wang2023scrimp}.

Since the goals are randomly initialized within allocation radius $R_{\textit{alloc}}$ around the agents' start locations, they can still be allocated in closer proximity to the agents which alleviates catastrophic forgetting of easier tasks that the agents have mastered before.

\section{Experimental Setup}\label{sec:experiments}

\subsection{Maps and Instances}\label{subsec:maps_instances}

The \emph{training maps} are randomly generated according to \cite{sartoretti2019primal} and have different shapes $K \times K$ defined by \emph{map size} $K \in \{10, 40, 80\}$. The \emph{obstacle density} $\delta \in \{0, 0.1, 0.2, 0.3\}$ defines the fraction of non-occupiable locations in the maps. All agents start at random locations with randomly assigned goals according to an allocation radius $R_{\textit{alloc}}$. If $R_{\textit{alloc}} = \infty$, then the goals can be placed anywhere on the training map.

The \emph{test maps} are provided by \cite{sartoretti2019primal}. For each map configuration of size $K$ and obstacle density $\delta$, there are 100 pre-generated test instances $I$ with fixed start and goal locations for all agents to ensure a fair comparison between different MARL approaches.

We always set $\gamma = 1$ as suggested in Section \ref{sec:mapf_stochastic_game}.

\subsection{Algorithms and Training}\label{subsec:marl_algorithms}

We implement PPO\footnote{Code available at \underline{\url{github.com/thomyphan/rl4mapf}}.} for policy learning according to Eq. \ref{eq:ppo_loss} and QMIX, QPLEX, and MAPPO for critic learning. We implement a purely RL-based version of PRIMAL using the same reward function as defined in \cite{sartoretti2019primal}. In addition, we employ a naive baseline, called \emph{No Curriculum}, only consisting of PPO and QMIX\footnote{We choose QMIX for consistency with our default setting. Replacing QMIX with QPLEX or MAPPO does not notably affect the performance of this baseline.} without any shaped reward. PRIMAL and \emph{No Curriculum} are trained with $R_{\textit{alloc}} = \infty$ (Section \ref{subsec:maps_instances}).

We train all algorithms on different training maps as explained in Section \ref{subsec:maps_instances} with $N = 8$ agents for 5000 epochs consisting of $E = 32$ episodes. The maps of size $K = 10$ are sampled twice as often as the other map sizes as proposed in \cite{sartoretti2019primal}. Each episode terminates after all agents reach their goal or after $T = 256$ time steps. All algorithms use parameter sharing, i.e., where all agents use the same policy and individual critic network $\hat{\pi}_{i}$ and $\hat{Q}_{i}$ respectively.

We denote \textit{CACTUS (X)} as CACTUS using MARL algorithm \textit{X} for critic learning. Unless stated otherwise, CACTUS always uses PPO for policy and \textit{X=QMIX} for critic learning.

In addition, we run \textit{CBSH} as a slow but optimal MAPF solver with a runtime limit of 5 minutes \cite{felner2018adding} and \textit{MAPF-LNS} \cite{li2021anytime} as a fast anytime MAPF solver with a runtime limit of 1 minute.

\subsection{Neural Networks and Hyperparameters}\label{subsec:neural_network_architectures}
For CACTUS and \textit{No Curriculum}, we use deep neural networks to implement $\hat{\pi}_{i}$ and $\hat{Q}_{i}$ for each agent $i$ and factorization operator $\Psi$ for QMIX and QPLEX. The neural networks are updated after every $E = 32$ episodes using ADAM with a learning rate of $0.001$.

Since all regarded maps are gridworlds, the observations are encoded as multi-channel image as illustrated in Fig. \ref{fig:observation_example}. We implement all neural networks as \emph{multilayer perceptron (MLP)} and flatten the multi-channel images before feeding them into the networks. $\hat{\pi}_{i}$ and $\hat{Q}_{i}$ have two hidden layers of 64 units with ELU activation. The output of $\hat{\pi}_{i}$ has $|\mathcal{A}_{i}|$ units with softmax activation. The output of $\hat{Q}_{i}$ has $|\mathcal{A}_{i}|$ linear units. The hypernetworks of QMIX as well as the critic of MAPPO have two hidden layers of 128 units with ELU activation and one or $|\mathcal{A}_{i}|$ linear output units respectively. For PRIMAL, we use the same architecture as proposed in \cite{sartoretti2019primal}.

Unless stated otherwise, CACTUS always uses a threshold of $U = 75\%$ and a deviation factor of $\eta = 2$, which corresponds to a confidence level of about 97\% in one-tailed tests.

\paragraph{\textbf{Computing Infrastructure}}

All training and test runs are performed on a x86\_64 GNU/Linux (Ubuntu 18.04.5 LTS) machine with i7-8700 @ 3.2GHz CPU (8 cores) and 64 GB RAM. Due to the simplicity of CACTUS, we do not need any GPU or distributed HPC infrastructure in contrast to \cite{sartoretti2019primal,damani2021primal,wang2023scrimp}.

\section{Results}\label{sec:results}

For each experiment, all respective algorithms are run 10 times to report the average progress and the 95\% confidence interval. We evaluate the training progress and generalization of trained policies with the pre-generated test instances $I$ explained in Section \ref{subsec:maps_instances}.

\subsection{Simplicity of CACTUS}\label{subsec:complexity_comparison}

To demonstrate the simplicity of CACTUS, we first quantify the training time, the training data w.r.t. the number of episodes, the number of trainable parameters, and the reward complexity and compare them with the original PRIMAL as specified in \cite{sartoretti2019primal}.

An overview is given in Table \ref{tab:simplicity_comparison}. In almost all aspects, CACTUS only requires 5\% or less of the effort of the original PRIMAL therefore being clearly the simpler and more efficient MARL approach to MAPF. Unlike PRIMAL, CACTUS is only run on CPU while still requiring significantly less training time. Furthermore, CACTUS does not depend on any expert data, i.e., recommendations of a centralized MAPF solver, which saves a significant amount of compute. While the reward function of PRIMAL requires four penalties for very specific situations, CACTUS is trained with a very simple reward function that penalizes any time step unless the goal is reached without considering any specific case (Section \ref{sec:mapf_stochastic_game}).

\begin{table}[!ht]
\centering
\caption{Comparison of the original PRIMAL and CACTUS w.r.t. various numbers in our experiments. The last column provides the amount of effort relative to the original PRIMAL. The numbers of the original PRIMAL are from \cite{sartoretti2019primal}. Unlike \cite{sartoretti2019primal}, we do not use any GPU or expert data for training.}
\begin{tabular}{|l|P{1.75cm}|c||P{1.25cm}|} \hline
 & PRIMAL (original \cite{sartoretti2019primal}) & CACTUS & Rel. to PRIMAL \\\hline
Training Time & $\approx 20$ days & $\approx 1$ day & $\approx 5\%$\\
\# Training Episodes & $\approx 3.8$ million & $160,000$ & $\approx 4.2\%$\\
\# Parameters & $\approx 13$ million & $579,979$ & $\approx 4.5\%$\\ 
\# Reward Penalties & 4 & 1 & $25\%$\\ \hline
\end{tabular}\label{tab:simplicity_comparison}
\end{table}

\begin{figure}[!ht]
	\centering
	\includegraphics[width=0.45\textwidth]{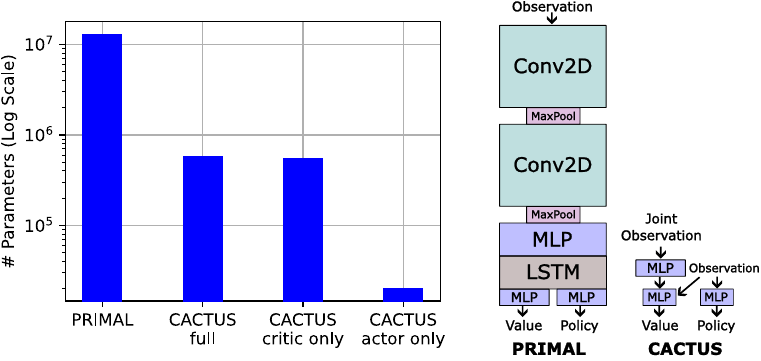}
     \caption{\textit{Left}: Comparison of the number of trainable parameters in PRIMAL and CACTUS. Note the logarithmic scale on the y-axis. \textit{Right}: The schematic network architectures used for PRIMAL and CACTUS. The sizes do not reflect any quantity and only illustrate the components used for learning.}
     \Description{\textit{Left}: Comparison of the number of trainable parameters in PRIMAL and CACTUS. Note the logarithmic scale on the y-axis. \textit{Right}: The schematic network architectures used for PRIMAL and CACTUS. The sizes do not reflect any quantity and only illustrate the components used for learning.}
     \label{fig:parameter_comparison}
\end{figure}

Fig. \ref{fig:parameter_comparison} compares the number of trainable parameters and neural network architectures of PRIMAL and CACTUS. In CACTUS, the critic with the mixing network has the majority of trainable parameters, which are only required during training. The actor size is negligible in CACTUS, which enables significantly faster inference than PRIMAL. The network architecture of CACTUS is also much simpler than PRIMAL since it is only based on MLPs thus does not depend on specialized hardware or significant computational effort for fine-tuning and training.

\subsection{Curriculum Learning}\label{subsec:curriculum_learning}

We evaluate the effect of CACTUS using QMIX, QPLEX, and MAPPO as current state-of-the-art MARL techniques \cite{rashid2018qmix,wang2020qplex,yu2022surprising}. After every epoch, we measure the average completion rate w.r.t. all test instances $I$ with map size $K \in \{10, 40, 80\}$ as well as obstacle density $\delta \in \{0, 0.1, 0.2, 0.3\}$ for $N = 8$ agents.

\begin{figure}
	\centering
	\includegraphics[width=0.45\textwidth]{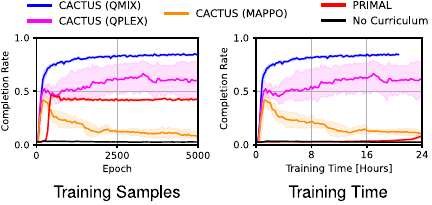}
     \caption{Average training progress of CACTUS variants, PRIMAL, and a naive MARL baseline without any curriculum w.r.t. training epochs (left) and training time (right). The performance is evaluated on all pre-generated test instances $I$ of \cite{sartoretti2019primal} with $K \in \{10, 40, 80\}$, $\delta \in \{0, 0.1, 0.2, 0.3\}$, and $N = 8$ agents. Shaded areas show the 95\% confidence interval.}
     \Description{Average training progress of CACTUS variants, PRIMAL, and a naive MARL baseline without any curriculum w.r.t. training epochs (left) and training time (right). The performance is evaluated on all pre-generated test instances $I$ of \cite{sartoretti2019primal} with $K \in \{10, 40, 80\}$, $\delta \in \{0, 0.1, 0.2, 0.3\}$, and $N = 8$ agents. Shaded areas show the 95\% confidence interval.}
     \label{fig:star_results}
\end{figure}

The results are shown in Fig. \ref{fig:star_results}. \textit{CACTUS (QMIX)} and \textit{CACTUS (QPLEX)} perform best. PRIMAL always outperforms \textit{CACTUS (MAPPO)}. \emph{No Curriculum} fails to learn any meaningful policy. However, PRIMAL is barely able to outperform \emph{No Curriculum} after 24 hours of training. \textit{CACTUS (QMIX)} completes training below 24 hours, while \textit{CACTUS (QPLEX)} and \textit{CACTUS (MAPPO)} require slightly more training time than 24 hours.

\subsection{CACTUS Hyperparameters}

Next, we evaluate the impact of different decision thresholds $U \in \{0.25, 0.5, 0.75\}$ and deviation factors $\eta = \{1, 2, 3\}$ on CACTUS. After every epoch, we measure the average completion rate w.r.t. all test instances $I$ with map size $K \in \{10, 40, 80\}$ as well as obstacle density $\delta \in \{0, 0.1, 0.2, 0.3\}$ for $N = 8$ agents. For the deviation factor evaluation, we consider CACTUS with $U = 0.25$, since all variants with $U = 0.75$ perform very similar as shown in Fig. \ref{fig:star_results}.

\begin{figure}[!ht]
	\centering
	\includegraphics[width=0.42\textwidth]{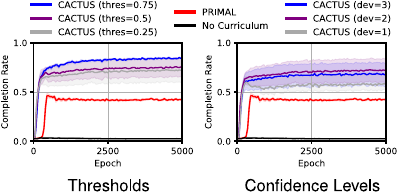}
     \caption{Average training progress of CACTUS variants, PRIMAL, and a naive MARL baseline without any curriculum w.r.t. different decision thresholds $U$ (left) and deviation factors $\eta$ (right). The performance is evaluated on all pre-generated test instances $I$ of \cite{sartoretti2019primal} with $K \in \{10, 40, 80\}$, $\delta \in \{0, 0.1, 0.2, 0.3\}$, and $N = 8$ agents. The right plot shows CACTUS variants with $U = 0.25$. Shaded areas show the 95\% confidence interval.}
     \Description{Average training progress of CACTUS variants, PRIMAL, and a naive MARL baseline without any curriculum w.r.t. different decision thresholds $U$ (left) and deviation factors $\eta$ (right). The performance is evaluated on all pre-generated test instances $I$ of \cite{sartoretti2019primal} with $K \in \{10, 40, 80\}$, $\delta \in \{0, 0.1, 0.2, 0.3\}$, and $N = 8$ agents. The right plot shows CACTUS variants with $U = 0.25$. Shaded areas show the 95\% confidence interval.}
     \label{fig:ablation_results}
\end{figure}

The results are shown in Fig. \ref{fig:ablation_results}. CACTUS performs best with $U = 0.75$ and second best with $U = 0.5$. CACTUS with $U = 0.25$ performs best when $\eta = 2$ and second best with $\eta = 3$. All CACTUS variants clearly outperform PRIMAL.

\subsection{Generalization}

Finally, we evaluate the generalization capabilities of policies trained with CACTUS using QMIX, QPLEX, and MAPPO as well as PRIMAL, and \textit{No Curriculum}. All policies are trained for 5000 epochs consisting of 32 episodes before being evaluated on all test instances $I$ with map size $K \in \{40, 80\}$ and obstacle density $\delta \in \{0, 0.2\}$ with different numbers of agents $N$. Note that all policies have only been trained with $N = 8$ (Section \ref{subsec:marl_algorithms}). We also report the average performance of the centralized MAPF solvers \textit{CBSH} and \textit{MAPF-LNS}.

\begin{figure}[!ht]
	\centering
	\includegraphics[width=0.45\textwidth]{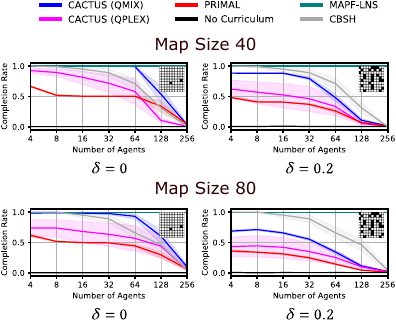}
     \caption{Average generalization performance of CACTUS variants as well as MARL and MAPF baselines to all test instances $I$ of size $K \in \{40, 80\}$ and obstacles density $\delta \in \{0, 0.1, 0.2, 0.3\}$ w.r.t. different numbers of agents $N$. The icons on the top-right of each plot show a $10 \times 10$ sub-grid example to illustrate the obstacle density. Shaded areas show the 95\% confidence interval.}
     \Description{Average generalization performance of CACTUS variants as well as MARL and MAPF baselines to all test instances $I$ of size $K \in \{40, 80\}$ and obstacles density $\delta \in \{0, 0.1, 0.2, 0.3\}$ w.r.t. different numbers of agents $N$. The icons on the top-right of each plot show a $10 \times 10$ sub-grid example to illustrate the obstacle density. Shaded areas show the 95\% confidence interval.}
     \label{fig:generalization_results}
\end{figure}

The generalization results w.r.t. different numbers of agents $N$ are shown in Fig. \ref{fig:generalization_results}. \textit{CACTUS (QMIX)} generalizes best compared to all other MARL approaches. In test instances with low obstacle density, \textit{CACTUS (QMIX)} always achieves an average completion rate over 70\% when scaling up to $N = 64$ agents. If $\delta = 0$, then \textit{CACTUS (QMIX)} is able to complete at least 50\% of all agents when scaling up to $N = 128$. \textit{CACTUS (QPLEX)} always outperforms PRIMAL on average except in instances with low obstacle density and map size $K = 40$, where the number of agents exceeds 32. PRIMAL is always outperformed by \textit{CACTUS (QMIX)} but consistently outperforms \textit{CACTUS (MAPPO)} and \emph{No Curriculum}. All approaches perform poorly when the agent number is $N \geq 256$ or $\delta = 0.2$.



Compared to the centralized MAPF solvers, \textit{CACTUS (QMIX)} can only outperform \textit{CBSH}, when the obstacle density $\delta$ is low. \textit{MAPF-LNS} is the best performing approach, always achieving a completion rate of 100\% in all test instances.

\subsection{Limitation in Structured Maps}

In addition, we tested CACTUS and the learning baselines in structured maps with rooms and narrow corridors such as mazes. While training was conducted on randomly generated maps as explained in Section \ref{subsec:maps_instances}, we evaluated the training progress using all maze and room maps of the common MAPF benchmark from \cite{stern2019multi}.

\begin{figure}
	\centering
	\includegraphics[width=0.45\textwidth]{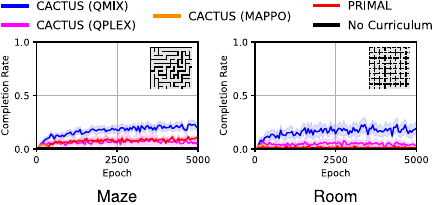}
     \caption{Average training progress of CACTUS variants, PRIMAL, and a naive MARL baseline without any curriculum in structured maps. The performance is evaluated on all maze and room maps provided by \cite{stern2019multi} respectively.}
     \Description{Average training progress of CACTUS variants, PRIMAL, and a naive MARL baseline without any curriculum in structured maps. The performance is evaluated on all maze and room maps provided by \cite{stern2019multi} respectively.}
     \label{fig:star_structure}
\end{figure}

The results are shown in Fig. \ref{fig:star_structure}. Compared to the results for unstructured maps in Section \ref{subsec:curriculum_learning}, all approaches perform significantly worse with none of them reaching an average completion rate above 25\%. \textit{CACTUS (QMIX)} consistently outperforms all other approaches, which generally fail to complete more than 10\% of all agent tasks.

\section{Discussion}\label{sec:discussion}

In this paper, we presented CACTUS as a lightweight MARL approach to MAPF. CACTUS defines a simple reverse curriculum scheme, where the goal of each agent is randomly placed within an allocation radius around the agent's start location. The allocation radius increases gradually as all agents improve, which is assessed by a confidence-based measure.

Our results confirm the necessity of adequate curricula, as standard MARL methods without any curriculum would likely fail to learn any meaningful policy in our MAPF setting (Section \ref{sec:mapf_stochastic_game}). This is due to the sparse reward and the dynamic constraints of the problem formulation. The shaped reward function of PRIMAL is helpful in improving performance over standard MARL. Due to the many specifically defined penalties, PRIMAL policies are rather conservative, which leads to a performance plateau that cannot be overcome without imitation learning on suitable expert data. However, CACTUS with QMIX or QPLEX clearly outperforms PRIMAL with 95\% less training time and learnable parameters without any additional reward shaping or expert data. CACTUS with MAPPO performs poorly, which indicates the importance of adequate credit assignment mechanisms, e.g, value factorization, to address the coordination problem in MAPF.

CACTUS is robust w.r.t. the choice of hyperparameters as any configuration with decision threshold $U \geq 25\%$ and deviation factor $\eta \geq 1$ outperforms PRIMAL as shown in Fig. \ref{fig:ablation_results}. However, the decision threshold $U$ should be sufficiently high (at least 50\%) to ensure an adequate difficulty level for the agents. Choosing a confidence level that is too high, e.g., using $\eta > 3$, could result in slow curriculum updates, where agents may overfit on easy tasks.

Despite the relatively restrictive time and data budget compared to the original PRIMAL, CACTUS generalizes quite well with value factorization over different numbers of agents $N$ and map sizes $K$. CACTUS with QMIX scales up to instances with 8 to 16 times more agents than used during training, in contrast to standard MARL, which generally fails to learn any meaningful policy in the MAPF problem. While generalizing better than the alternative approaches, CACTUS still has some limitations regarding high obstacle density, large map sizes and structured maps with separate rooms and narrow corridors, where there is still potential for improvement. Despite the ability of efficient decentralized decision-making, CACTUS is not competitive to centralized MAPF solvers due to its limited scalability.

Nevertheless, CACTUS clearly demonstrates how simple and well-defined curricula can enhance MARL techniques for MAPF without relying on extremely large neural networks, extensively shaped reward functions, or centralized MAPF solvers for imitation learning. Therefore, we hope to provide a suitable foundation to enable faster progress in connecting the areas of MARL and MAPF with significantly lower costs.






\section*{Acknowledgements}
The research at the University of Southern California was supported by the National Science Foundation (NSF) under grant numbers 1817189, 1837779, 1935712, 2121028, 2112533, and 2321786 as well as a gift from Amazon Robotics. The research at the Georgia Institute of Technology was supported by NSF under grant number 2112533. The views and conclusions contained in this document are those of the authors and should not be interpreted as representing the official policies, either expressed or implied, of the sponsoring organizations, agencies, or the U.S. government.

\balance
\bibliographystyle{ACM-Reference-Format} 
\bibliography{references}


\begin{thebibliography}{52}


\ifx \showCODEN    \undefined \def \showCODEN     #1{\unskip}     \fi
\ifx \showDOI      \undefined \def \showDOI       #1{#1}\fi
\ifx \showISBNx    \undefined \def \showISBNx     #1{\unskip}     \fi
\ifx \showISBNxiii \undefined \def \showISBNxiii  #1{\unskip}     \fi
\ifx \showISSN     \undefined \def \showISSN      #1{\unskip}     \fi
\ifx \showLCCN     \undefined \def \showLCCN      #1{\unskip}     \fi
\ifx \shownote     \undefined \def \shownote      #1{#1}          \fi
\ifx \showarticletitle \undefined \def \showarticletitle #1{#1}   \fi
\ifx \showURL      \undefined \def \showURL       {\relax}        \fi
\providecommand\bibfield[2]{#2}
\providecommand\bibinfo[2]{#2}
\providecommand\natexlab[1]{#1}
\providecommand\showeprint[2][]{arXiv:#2}

\bibitem[\protect\citeauthoryear{Agostinelli, McAleer, Shmakov, and
  Baldi}{Agostinelli et~al\mbox{.}}{2019}]%
        {agostinelli2019solving}
\bibfield{author}{\bibinfo{person}{Forest Agostinelli},
  \bibinfo{person}{Stephen McAleer}, \bibinfo{person}{Alexander Shmakov}, {and}
  \bibinfo{person}{Pierre Baldi}.} \bibinfo{year}{2019}\natexlab{}.
\newblock \showarticletitle{{Solving} the {Rubik}’s {Cube} with {Deep}
  {Reinforcement} {Learning} and {Search}}.
\newblock \bibinfo{journal}{\emph{Nature Machine Intelligence}}
  \bibinfo{volume}{1}, \bibinfo{number}{8} (\bibinfo{year}{2019}),
  \bibinfo{pages}{356--363}.
\newblock


\bibitem[\protect\citeauthoryear{Asada, Noda, Tawaratsumida, and Hosoda}{Asada
  et~al\mbox{.}}{1996}]%
        {asada1996purposive}
\bibfield{author}{\bibinfo{person}{Minoru Asada}, \bibinfo{person}{Shoichi
  Noda}, \bibinfo{person}{Sukoya Tawaratsumida}, {and} \bibinfo{person}{Koh
  Hosoda}.} \bibinfo{year}{1996}\natexlab{}.
\newblock \showarticletitle{{Purposive} {Behavior} {Acquisition} for a {Real}
  {Robot} by {Vision}-{Based} {Reinforcement} {Learning}}.
\newblock \bibinfo{journal}{\emph{Machine learning}}  \bibinfo{volume}{23}
  (\bibinfo{year}{1996}), \bibinfo{pages}{279--303}.
\newblock


\bibitem[\protect\citeauthoryear{Bengio, Louradour, Collobert, and
  Weston}{Bengio et~al\mbox{.}}{2009}]%
        {bengio2009curriculum}
\bibfield{author}{\bibinfo{person}{Yoshua Bengio},
  \bibinfo{person}{J{\'e}r{\^o}me Louradour}, \bibinfo{person}{Ronan
  Collobert}, {and} \bibinfo{person}{Jason Weston}.}
  \bibinfo{year}{2009}\natexlab{}.
\newblock \showarticletitle{{Curriculum} {Learning}}. In
  \bibinfo{booktitle}{\emph{26th International Conference on Machine
  Learning}}.
\newblock


\bibitem[\protect\citeauthoryear{Bu{\c{s}}oniu, Babu{\v{s}}ka, and
  De~Schutter}{Bu{\c{s}}oniu et~al\mbox{.}}{2010}]%
        {bucsoniu2010multi}
\bibfield{author}{\bibinfo{person}{Lucian Bu{\c{s}}oniu},
  \bibinfo{person}{Robert Babu{\v{s}}ka}, {and} \bibinfo{person}{Bart
  De~Schutter}.} \bibinfo{year}{2010}\natexlab{}.
\newblock \showarticletitle{{Multi}-{Agent} {Reinforcement} {Learning}: {An}
  {Overview}}.
\newblock \bibinfo{journal}{\emph{Innovations in Multi-Agent Systems and
  Applications-1}} (\bibinfo{year}{2010}), \bibinfo{pages}{183--221}.
\newblock


\bibitem[\protect\citeauthoryear{Cohen and Koenig}{Cohen and Koenig}{2016}]%
        {cohen2016bounded}
\bibfield{author}{\bibinfo{person}{Liron Cohen} {and} \bibinfo{person}{Sven
  Koenig}.} \bibinfo{year}{2016}\natexlab{}.
\newblock \showarticletitle{{Bounded} {Suboptimal} {Multi}-{Agent} {Path}
  {Finding} {Using} {Highways}}. In \bibinfo{booktitle}{\emph{IJCAI}}.
  \bibinfo{pages}{3978--3979}.
\newblock


\bibitem[\protect\citeauthoryear{Damani, Luo, Wenzel, and Sartoretti}{Damani
  et~al\mbox{.}}{2021}]%
        {damani2021primal}
\bibfield{author}{\bibinfo{person}{Mehul Damani}, \bibinfo{person}{Zhiyao Luo},
  \bibinfo{person}{Emerson Wenzel}, {and} \bibinfo{person}{Guillaume
  Sartoretti}.} \bibinfo{year}{2021}\natexlab{}.
\newblock \showarticletitle{{PRIMAL} $ \_2 $: {Pathfinding} via {Reinforcement}
  and {Imitation} {Multi}-{Agent} {Learning}-{Lifelong}}.
\newblock \bibinfo{journal}{\emph{IEEE Robotics and Automation Letters}}
  \bibinfo{volume}{6}, \bibinfo{number}{2} (\bibinfo{year}{2021}),
  \bibinfo{pages}{2666--2673}.
\newblock


\bibitem[\protect\citeauthoryear{Dennis, Jaques, Vinitsky, Bayen, Russell,
  Critch, and Levine}{Dennis et~al\mbox{.}}{2020}]%
        {dennis2020emergent}
\bibfield{author}{\bibinfo{person}{Michael Dennis}, \bibinfo{person}{Natasha
  Jaques}, \bibinfo{person}{Eugene Vinitsky}, \bibinfo{person}{Alexandre
  Bayen}, \bibinfo{person}{Stuart Russell}, \bibinfo{person}{Andrew Critch},
  {and} \bibinfo{person}{Sergey Levine}.} \bibinfo{year}{2020}\natexlab{}.
\newblock \showarticletitle{{Emergent} {Complexity} and {Zero}-{Shot}
  {Transfer} via {Unsupervised} {Environment} {Design}}.
\newblock \bibinfo{journal}{\emph{NeurIPS}}  \bibinfo{volume}{33}
  (\bibinfo{year}{2020}).
\newblock


\bibitem[\protect\citeauthoryear{Du, Abbeel, and Grover}{Du
  et~al\mbox{.}}{2021}]%
        {du2021takes}
\bibfield{author}{\bibinfo{person}{Yuqing Du}, \bibinfo{person}{Pieter Abbeel},
  {and} \bibinfo{person}{Aditya Grover}.} \bibinfo{year}{2021}\natexlab{}.
\newblock \showarticletitle{{It} {Takes} {Four} to {Tango}: {Multiagent} {Self}
  {Play} for {Automatic} {Curriculum} {Generation}}. In
  \bibinfo{booktitle}{\emph{ICLR}}.
\newblock


\bibitem[\protect\citeauthoryear{Felner, Li, Boyarski, Ma, Cohen, Kumar, and
  Koenig}{Felner et~al\mbox{.}}{2018}]%
        {felner2018adding}
\bibfield{author}{\bibinfo{person}{Ariel Felner}, \bibinfo{person}{Jiaoyang
  Li}, \bibinfo{person}{Eli Boyarski}, \bibinfo{person}{Hang Ma},
  \bibinfo{person}{Liron Cohen}, \bibinfo{person}{TK~Satish Kumar}, {and}
  \bibinfo{person}{Sven Koenig}.} \bibinfo{year}{2018}\natexlab{}.
\newblock \showarticletitle{{Adding} {Heuristics} to {Conflict}-{Based}
  {Search} for {Multi}-{Agent} {Path} {Finding}}. In
  \bibinfo{booktitle}{\emph{ICAPS}}, Vol.~\bibinfo{volume}{28}.
  \bibinfo{pages}{83--87}.
\newblock


\bibitem[\protect\citeauthoryear{Florensa, Held, Wulfmeier, Zhang, and
  Abbeel}{Florensa et~al\mbox{.}}{2017}]%
        {florensa2017reverse}
\bibfield{author}{\bibinfo{person}{Carlos Florensa}, \bibinfo{person}{David
  Held}, \bibinfo{person}{Markus Wulfmeier}, \bibinfo{person}{Michael Zhang},
  {and} \bibinfo{person}{Pieter Abbeel}.} \bibinfo{year}{2017}\natexlab{}.
\newblock \showarticletitle{{Reverse} {Curriculum} {Generation} for
  {Reinforcement} {Learning}}. In \bibinfo{booktitle}{\emph{Conference on Robot
  Learning}}. PMLR, \bibinfo{pages}{482--495}.
\newblock


\bibitem[\protect\citeauthoryear{Gabor, Sedlmeier, Kiermeier, Phan,
  et~al\mbox{.}}{Gabor et~al\mbox{.}}{2019}]%
        {gabor2019scenario}
\bibfield{author}{\bibinfo{person}{Thomas Gabor}, \bibinfo{person}{Andreas
  Sedlmeier}, \bibinfo{person}{Marie Kiermeier}, \bibinfo{person}{Thomy Phan},
  {et~al\mbox{.}}} \bibinfo{year}{2019}\natexlab{}.
\newblock \showarticletitle{{Scenario} {Co}-{Evolution} for {Reinforcement}
  {Learning} on a {Grid} {World} {Smart} {Factory} {Domain}}. In
  \bibinfo{booktitle}{\emph{Genetic and Evolutionary Computation Conference}}.
  \bibinfo{pages}{898–906}.
\newblock
\showISBNx{9781450361118}


\bibitem[\protect\citeauthoryear{Huang, Koenig, and Dilkina}{Huang
  et~al\mbox{.}}{2021}]%
        {huang2021learning}
\bibfield{author}{\bibinfo{person}{Taoan Huang}, \bibinfo{person}{Sven Koenig},
  {and} \bibinfo{person}{Bistra Dilkina}.} \bibinfo{year}{2021}\natexlab{}.
\newblock \showarticletitle{{Learning} to {Resolve} {Conflicts} for
  {Multi}-{Agent} {Path} {Finding} with {Conflict}-{Based} {Search}}. In
  \bibinfo{booktitle}{\emph{AAAI Conference on Artificial Intelligence}},
  Vol.~\bibinfo{volume}{35}. \bibinfo{pages}{11246--11253}.
\newblock


\bibitem[\protect\citeauthoryear{Huang, Li, Koenig, and Dilkina}{Huang
  et~al\mbox{.}}{2022}]%
        {HuangAAAI22}
\bibfield{author}{\bibinfo{person}{Taoan Huang}, \bibinfo{person}{Jiaoyang Li},
  \bibinfo{person}{Sven Koenig}, {and} \bibinfo{person}{Bistra Dilkina}.}
  \bibinfo{year}{2022}\natexlab{}.
\newblock \showarticletitle{{Anytime} {Multi}-{Agent} {Path} {Finding} via
  {Machine} {Learning}-{Guided} {Large} {Neighborhood} {Search}}. In
  \bibinfo{booktitle}{\emph{36th AAAI Conference on Artificial Intelligence
  (AAAI)}}. \bibinfo{pages}{9368--9376}.
\newblock


\bibitem[\protect\citeauthoryear{Jaderberg, Czarnecki, Dunning, Marris, Lever,
  et~al\mbox{.}}{Jaderberg et~al\mbox{.}}{2019}]%
        {jaderberg2019human}
\bibfield{author}{\bibinfo{person}{Max Jaderberg}, \bibinfo{person}{Wojciech~M
  Czarnecki}, \bibinfo{person}{Iain Dunning}, \bibinfo{person}{Luke Marris},
  \bibinfo{person}{Lever}, {et~al\mbox{.}}} \bibinfo{year}{2019}\natexlab{}.
\newblock \showarticletitle{{Human}-{Level} {Performance} in {3D} {Multiplayer}
  {Games} with {Population}-based {Reinforcement} {Learning}}.
\newblock \bibinfo{journal}{\emph{Science}} \bibinfo{volume}{364},
  \bibinfo{number}{6443} (\bibinfo{year}{2019}).
\newblock


\bibitem[\protect\citeauthoryear{Kaduri, Boyarski, and Stern}{Kaduri
  et~al\mbox{.}}{2020}]%
        {kaduri2020algorithm}
\bibfield{author}{\bibinfo{person}{Omri Kaduri}, \bibinfo{person}{Eli
  Boyarski}, {and} \bibinfo{person}{Roni Stern}.}
  \bibinfo{year}{2020}\natexlab{}.
\newblock \showarticletitle{{Algorithm} {Selection} for {Optimal}
  {Multi}-{Agent} {Pathfinding}}. In \bibinfo{booktitle}{\emph{ICAPS}},
  Vol.~\bibinfo{volume}{30}. \bibinfo{pages}{161--165}.
\newblock


\bibitem[\protect\citeauthoryear{Li, Chen, Harabor, Stuckey, and Koenig}{Li
  et~al\mbox{.}}{2021a}]%
        {li2021anytime}
\bibfield{author}{\bibinfo{person}{Jiaoyang Li}, \bibinfo{person}{Zhe Chen},
  \bibinfo{person}{Daniel Harabor}, \bibinfo{person}{Peter~J. Stuckey}, {and}
  \bibinfo{person}{Sven Koenig}.} \bibinfo{year}{2021}\natexlab{a}.
\newblock \showarticletitle{{Anytime} {Multi}-{Agent} {Path} {Finding} via
  {Large} {Neighborhood} {Search}}. In \bibinfo{booktitle}{\emph{International
  Joint Conference on Artificial Intelligence (IJCAI)}}.
  \bibinfo{pages}{4127--4135}.
\newblock


\bibitem[\protect\citeauthoryear{Li, Tinka, Kiesel, Durham, Kumar, and
  Koenig}{Li et~al\mbox{.}}{2021b}]%
        {li2021lifelong}
\bibfield{author}{\bibinfo{person}{Jiaoyang Li}, \bibinfo{person}{Andrew
  Tinka}, \bibinfo{person}{Scott Kiesel}, \bibinfo{person}{Joseph~W Durham},
  \bibinfo{person}{TK~Satish Kumar}, {and} \bibinfo{person}{Sven Koenig}.}
  \bibinfo{year}{2021}\natexlab{b}.
\newblock \showarticletitle{{Lifelong} {Multi}-{Agent} {Path} {Finding} in
  {Large}-{Scale} {Warehouses}}. In \bibinfo{booktitle}{\emph{AAAI Conference
  on Artificial Intelligence}}, Vol.~\bibinfo{volume}{35}.
  \bibinfo{pages}{11272--11281}.
\newblock


\bibitem[\protect\citeauthoryear{Lowe, Wu, Tamar, Harb, Abbeel, and
  Mordatch}{Lowe et~al\mbox{.}}{2017}]%
        {lowe2017multi}
\bibfield{author}{\bibinfo{person}{Ryan Lowe}, \bibinfo{person}{Yi Wu},
  \bibinfo{person}{Aviv Tamar}, \bibinfo{person}{Jean Harb},
  \bibinfo{person}{Pieter Abbeel}, {and} \bibinfo{person}{Igor Mordatch}.}
  \bibinfo{year}{2017}\natexlab{}.
\newblock \showarticletitle{{Multi}-{Agent} {Actor}-{Critic} for {Mixed}
  {Cooperative}-{Competitive} {Environments}}. In
  \bibinfo{booktitle}{\emph{Advances in Neural Information Processing
  Systems}}. \bibinfo{pages}{6379--6390}.
\newblock


\bibitem[\protect\citeauthoryear{McAleer, Agostinelli, Shmakov, and
  Baldi}{McAleer et~al\mbox{.}}{2018}]%
        {mcaleer2018solving}
\bibfield{author}{\bibinfo{person}{Stephen McAleer}, \bibinfo{person}{Forest
  Agostinelli}, \bibinfo{person}{Alexander Shmakov}, {and}
  \bibinfo{person}{Pierre Baldi}.} \bibinfo{year}{2018}\natexlab{}.
\newblock \showarticletitle{{Solving} the {Rubik}'s {Cube} with {Approximate}
  {Policy} {Iteration}}. In \bibinfo{booktitle}{\emph{ICLR}}.
\newblock


\bibitem[\protect\citeauthoryear{Mnih, Kavukcuoglu, Silver, Rusu,
  et~al\mbox{.}}{Mnih et~al\mbox{.}}{2015}]%
        {mnih2015human}
\bibfield{author}{\bibinfo{person}{Volodymyr Mnih}, \bibinfo{person}{Koray
  Kavukcuoglu}, \bibinfo{person}{David Silver}, \bibinfo{person}{Rusu},
  {et~al\mbox{.}}} \bibinfo{year}{2015}\natexlab{}.
\newblock \showarticletitle{{Human}-{Level} {Control} through {Deep}
  {Reinforcement} {Learning}}.
\newblock \bibinfo{journal}{\emph{Nature}} \bibinfo{volume}{518},
  \bibinfo{number}{7540} (\bibinfo{year}{2015}).
\newblock


\bibitem[\protect\citeauthoryear{Narvekar, Peng, Leonetti, Sinapov, Taylor, and
  Stone}{Narvekar et~al\mbox{.}}{2020}]%
        {narvekar2020curriculum}
\bibfield{author}{\bibinfo{person}{Sanmit Narvekar}, \bibinfo{person}{Bei
  Peng}, \bibinfo{person}{Matteo Leonetti}, \bibinfo{person}{Jivko Sinapov},
  \bibinfo{person}{Matthew~E Taylor}, {and} \bibinfo{person}{Peter Stone}.}
  \bibinfo{year}{2020}\natexlab{}.
\newblock \showarticletitle{{Curriculum} {Learning} for {Reinforcement}
  {Learning} {Domains}: {A} {Framework} and {Survey}}.
\newblock \bibinfo{journal}{\emph{The Journal of Machine Learning Research}}
  \bibinfo{volume}{21}, \bibinfo{number}{1} (\bibinfo{year}{2020}).
\newblock


\bibitem[\protect\citeauthoryear{Narvekar, Sinapov, Leonetti, and
  Stone}{Narvekar et~al\mbox{.}}{2016}]%
        {narvekar2016source}
\bibfield{author}{\bibinfo{person}{Sanmit Narvekar}, \bibinfo{person}{Jivko
  Sinapov}, \bibinfo{person}{Matteo Leonetti}, {and} \bibinfo{person}{Peter
  Stone}.} \bibinfo{year}{2016}\natexlab{}.
\newblock \showarticletitle{{Source} {Task} {Creation} for {Curriculum}
  {Learning}}. In \bibinfo{booktitle}{\emph{AAMAS}}.
\newblock


\bibitem[\protect\citeauthoryear{Pham and Bera}{Pham and Bera}{2023}]%
        {pham2023crowd}
\bibfield{author}{\bibinfo{person}{Phu Pham} {and} \bibinfo{person}{Aniket
  Bera}.} \bibinfo{year}{2023}\natexlab{}.
\newblock \showarticletitle{{Crowd}-{Aware} {Multi}-{Agent} {Pathfinding} with
  {Boosted} {Curriculum} {Reinforcement} {Learning}}.
\newblock \bibinfo{journal}{\emph{arXiv preprint arXiv:2309.10275}}
  (\bibinfo{year}{2023}).
\newblock


\bibitem[\protect\citeauthoryear{Phan, Belzner, Gabor, Sedlmeier, Ritz, and
  Linnhoff-Popien}{Phan et~al\mbox{.}}{2021a}]%
        {phan2021resilient}
\bibfield{author}{\bibinfo{person}{Thomy Phan}, \bibinfo{person}{Lenz Belzner},
  \bibinfo{person}{Thomas Gabor}, \bibinfo{person}{Andreas Sedlmeier},
  \bibinfo{person}{Fabian Ritz}, {and} \bibinfo{person}{Claudia
  Linnhoff-Popien}.} \bibinfo{year}{2021}\natexlab{a}.
\newblock \showarticletitle{{Resilient} {Multi}-{Agent} {Reinforcement}
  {Learning} with {Adversarial} {Value} {Decomposition}}.
\newblock \bibinfo{journal}{\emph{AAAI Conference on Artificial Intelligence}}
  \bibinfo{number}{13} (\bibinfo{year}{2021}).
\newblock


\bibitem[\protect\citeauthoryear{Phan, Huang, Dilkina, and Koenig}{Phan
  et~al\mbox{.}}{2024}]%
        {phanAAAI24}
\bibfield{author}{\bibinfo{person}{Thomy Phan}, \bibinfo{person}{Taoan Huang},
  \bibinfo{person}{Bistra Dilkina}, {and} \bibinfo{person}{Sven Koenig}.}
  \bibinfo{year}{2024}\natexlab{}.
\newblock \showarticletitle{{Adaptive} {Anytime} {Multi}-{Agent} {Path}
  {Finding} {Using} {Bandit}-{Based} {Large} {Neighborhood} {Search}}.
\newblock \bibinfo{journal}{\emph{Proceedings of the AAAI Conference on
  Artificial Intelligence (AAAI)}} (\bibinfo{year}{2024}).
\newblock
\urldef\tempurl%
\url{https://thomyphan.github.io/publication/2024-02-01-aaai-phan}
\showURL{%
\tempurl}


\bibitem[\protect\citeauthoryear{Phan, Ritz, Altmann, Zorn, N\"{u}{\ss}lein,
  K\"{o}lle, Gabor, and Linnhoff-Popien}{Phan et~al\mbox{.}}{2023}]%
        {phan2023attention}
\bibfield{author}{\bibinfo{person}{Thomy Phan}, \bibinfo{person}{Fabian Ritz},
  \bibinfo{person}{Philipp Altmann}, \bibinfo{person}{Maximilian Zorn},
  \bibinfo{person}{Jonas N\"{u}{\ss}lein}, \bibinfo{person}{Michael K\"{o}lle},
  \bibinfo{person}{Thomas Gabor}, {and} \bibinfo{person}{Claudia
  Linnhoff-Popien}.} \bibinfo{year}{2023}\natexlab{}.
\newblock \showarticletitle{{Attention}-{Based} {Recurrence} for
  {Multi}-{Agent} {Reinforcement} {Learning} under {Stochastic} {Partial}
  {Observability}}. In \bibinfo{booktitle}{\emph{40th International Conference
  on Machine Learning}}.
\newblock


\bibitem[\protect\citeauthoryear{Phan, Ritz, Belzner, Altmann, Gabor, and
  Linnhoff-Popien}{Phan et~al\mbox{.}}{2021b}]%
        {phan2021vast}
\bibfield{author}{\bibinfo{person}{Thomy Phan}, \bibinfo{person}{Fabian Ritz},
  \bibinfo{person}{Lenz Belzner}, \bibinfo{person}{Philipp Altmann},
  \bibinfo{person}{Thomas Gabor}, {and} \bibinfo{person}{Claudia
  Linnhoff-Popien}.} \bibinfo{year}{2021}\natexlab{b}.
\newblock \showarticletitle{{VAST}: {Value} {Function} {Factorization} with
  {Variable} {Agent} {Sub}-{Teams}}. In \bibinfo{booktitle}{\emph{Advances in
  Neural Information Processing Systems}}, Vol.~\bibinfo{volume}{34}.
  \bibinfo{publisher}{Curran Associates, Inc.}, \bibinfo{pages}{24018--24032}.
\newblock


\bibitem[\protect\citeauthoryear{Rashid, Samvelyan, de~Witt, Farquhar,
  Foerster, and Whiteson}{Rashid et~al\mbox{.}}{2018}]%
        {rashid2018qmix}
\bibfield{author}{\bibinfo{person}{Tabish Rashid}, \bibinfo{person}{Mikayel
  Samvelyan}, \bibinfo{person}{Christian~Schroeder de Witt},
  \bibinfo{person}{Gregory Farquhar}, \bibinfo{person}{Jakob Foerster}, {and}
  \bibinfo{person}{Shimon Whiteson}.} \bibinfo{year}{2018}\natexlab{}.
\newblock \showarticletitle{{QMIX}: {Monotonic} {Value} {Function}
  {Factorisation} for {Deep} {Multi}-{Agent} {Reinforcement} {Learning}}. In
  \bibinfo{booktitle}{\emph{35th International Conference on Machine
  Learning}}. \bibinfo{publisher}{PMLR}, \bibinfo{pages}{4295--4304}.
\newblock


\bibitem[\protect\citeauthoryear{Ratner and Warmuth}{Ratner and
  Warmuth}{1986}]%
        {ratner1986finding}
\bibfield{author}{\bibinfo{person}{Daniel Ratner} {and}
  \bibinfo{person}{Manfred Warmuth}.} \bibinfo{year}{1986}\natexlab{}.
\newblock \showarticletitle{{Finding} a {Shortest} {Solution} for the {N}x{N}
  {Extension} of the 15-{Puzzle} is {Intractable}}. In
  \bibinfo{booktitle}{\emph{5th AAAI National Conference on Artificial
  Intelligence}}. \bibinfo{publisher}{AAAI Press}, \bibinfo{pages}{168–172}.
\newblock


\bibitem[\protect\citeauthoryear{Sartoretti, Kerr, Shi, Wagner, Kumar,
  et~al\mbox{.}}{Sartoretti et~al\mbox{.}}{2019}]%
        {sartoretti2019primal}
\bibfield{author}{\bibinfo{person}{Guillaume Sartoretti},
  \bibinfo{person}{Justin Kerr}, \bibinfo{person}{Yunfei Shi},
  \bibinfo{person}{Glenn Wagner}, \bibinfo{person}{TK~Satish Kumar},
  {et~al\mbox{.}}} \bibinfo{year}{2019}\natexlab{}.
\newblock \showarticletitle{{PRIMAL}: {Pathfinding} via {Reinforcement} and
  {Imitation} {Multi}-{Agent} {Learning}}.
\newblock \bibinfo{journal}{\emph{IEEE Robotics and Automation Letters}}
  \bibinfo{volume}{4}, \bibinfo{number}{3} (\bibinfo{year}{2019}),
  \bibinfo{pages}{2378--2385}.
\newblock


\bibitem[\protect\citeauthoryear{Schulman, Wolski, Dhariwal,
  et~al\mbox{.}}{Schulman et~al\mbox{.}}{2017}]%
        {schulman2017proximal}
\bibfield{author}{\bibinfo{person}{John Schulman}, \bibinfo{person}{Filip
  Wolski}, \bibinfo{person}{Prafulla Dhariwal}, {et~al\mbox{.}}}
  \bibinfo{year}{2017}\natexlab{}.
\newblock \showarticletitle{{Proximal} {Policy} {Optimization} {Algorithms}}.
\newblock \bibinfo{journal}{\emph{arXiv preprint arXiv:1707.06347}}
  (\bibinfo{year}{2017}).
\newblock


\bibitem[\protect\citeauthoryear{Sharon, Stern, Felner, and Sturtevant}{Sharon
  et~al\mbox{.}}{2012}]%
        {sharon2012conflict}
\bibfield{author}{\bibinfo{person}{Guni Sharon}, \bibinfo{person}{Roni Stern},
  \bibinfo{person}{Ariel Felner}, {and} \bibinfo{person}{Nathan Sturtevant}.}
  \bibinfo{year}{2012}\natexlab{}.
\newblock \showarticletitle{{Conflict}-{Based} {Search} {For} {Optimal}
  {Multi}-{Agent} {Path} {Finding}}.
\newblock \bibinfo{journal}{\emph{AAAI Conference on Artificial Intelligence}}
  \bibinfo{volume}{26}, \bibinfo{number}{1} (\bibinfo{date}{Sep.}
  \bibinfo{year}{2012}), \bibinfo{pages}{563--569}.
\newblock


\bibitem[\protect\citeauthoryear{Silva and Costa}{Silva and Costa}{2018}]%
        {silva2018object}
\bibfield{author}{\bibinfo{person}{Felipe Leno~Da Silva} {and}
  \bibinfo{person}{Anna Helena~Reali Costa}.} \bibinfo{year}{2018}\natexlab{}.
\newblock \showarticletitle{{Object}-{Oriented} {Curriculum} {Generation} for
  {Reinforcement} {Learning}}. In \bibinfo{booktitle}{\emph{17th International
  Conference on Autonomous Agents and Multiagent Systems}}.
  \bibinfo{pages}{1026--1034}.
\newblock


\bibitem[\protect\citeauthoryear{Silver}{Silver}{2005}]%
        {silver2005cooperative}
\bibfield{author}{\bibinfo{person}{David Silver}.}
  \bibinfo{year}{2005}\natexlab{}.
\newblock \showarticletitle{{Cooperative} {Pathfinding}}.
\newblock \bibinfo{journal}{\emph{AAAI Conference on Artificial Intelligence
  and Interactive Digital Entertainment}} \bibinfo{volume}{1},
  \bibinfo{number}{1} (\bibinfo{date}{Sep.} \bibinfo{year}{2005}),
  \bibinfo{pages}{117--122}.
\newblock


\bibitem[\protect\citeauthoryear{Silver, Schrittwieser, Simonyan, Antonoglou,
  et~al\mbox{.}}{Silver et~al\mbox{.}}{2017}]%
        {silver2017mastering}
\bibfield{author}{\bibinfo{person}{David Silver}, \bibinfo{person}{Julian
  Schrittwieser}, \bibinfo{person}{Karen Simonyan}, \bibinfo{person}{Ioannis
  Antonoglou}, {et~al\mbox{.}}} \bibinfo{year}{2017}\natexlab{}.
\newblock \showarticletitle{{Mastering} the {Game} of {Go} without {Human}
  {Knowledge}}.
\newblock \bibinfo{journal}{\emph{Nature}} \bibinfo{volume}{550},
  \bibinfo{number}{7676} (\bibinfo{year}{2017}), \bibinfo{pages}{354--359}.
\newblock


\bibitem[\protect\citeauthoryear{Skalse, Howe, Krasheninnikov, and
  Krueger}{Skalse et~al\mbox{.}}{2022}]%
        {skalse2022defining}
\bibfield{author}{\bibinfo{person}{Joar Skalse}, \bibinfo{person}{Nikolaus
  Howe}, \bibinfo{person}{Dmitrii Krasheninnikov}, {and} \bibinfo{person}{David
  Krueger}.} \bibinfo{year}{2022}\natexlab{}.
\newblock \showarticletitle{{Defining} and {Characterizing} {Reward} {Gaming}}.
\newblock \bibinfo{journal}{\emph{Advances in Neural Information Processing
  Systems}}  \bibinfo{volume}{35} (\bibinfo{year}{2022}),
  \bibinfo{pages}{9460--9471}.
\newblock


\bibitem[\protect\citeauthoryear{Son, Kim, Kang, Hostallero, and Yi}{Son
  et~al\mbox{.}}{2019}]%
        {son2019qtran}
\bibfield{author}{\bibinfo{person}{Kyunghwan Son}, \bibinfo{person}{Daewoo
  Kim}, \bibinfo{person}{Wan~Ju Kang}, \bibinfo{person}{David~Earl Hostallero},
  {and} \bibinfo{person}{Yung Yi}.} \bibinfo{year}{2019}\natexlab{}.
\newblock \showarticletitle{{QTRAN}: {Learning} to {Factorize} with
  {Transformation} for {Cooperative} {Multi}-{Agent} {Reinforcement}
  {Learning}}. In \bibinfo{booktitle}{\emph{36th International Conference on
  Machine Learning}}. \bibinfo{publisher}{PMLR}, \bibinfo{pages}{5887--5896}.
\newblock


\bibitem[\protect\citeauthoryear{Soviany, Ionescu, Rota, and Sebe}{Soviany
  et~al\mbox{.}}{2022}]%
        {soviany2022curriculum}
\bibfield{author}{\bibinfo{person}{Petru Soviany}, \bibinfo{person}{Radu~Tudor
  Ionescu}, \bibinfo{person}{Paolo Rota}, {and} \bibinfo{person}{Nicu Sebe}.}
  \bibinfo{year}{2022}\natexlab{}.
\newblock \showarticletitle{{Curriculum} {Learning}: {A} {Survey}}.
\newblock \bibinfo{journal}{\emph{International Journal of Computer Vision}}
  \bibinfo{volume}{130}, \bibinfo{number}{6} (\bibinfo{year}{2022}).
\newblock


\bibitem[\protect\citeauthoryear{Stern, Sturtevant, Felner, Koenig, Ma, Walker,
  Li, Atzmon, Cohen, Kumar, et~al\mbox{.}}{Stern et~al\mbox{.}}{2019}]%
        {stern2019multi}
\bibfield{author}{\bibinfo{person}{Roni Stern}, \bibinfo{person}{Nathan
  Sturtevant}, \bibinfo{person}{Ariel Felner}, \bibinfo{person}{Sven Koenig},
  \bibinfo{person}{Hang Ma}, \bibinfo{person}{Thayne Walker},
  \bibinfo{person}{Jiaoyang Li}, \bibinfo{person}{Dor Atzmon},
  \bibinfo{person}{Liron Cohen}, \bibinfo{person}{TK Kumar}, {et~al\mbox{.}}}
  \bibinfo{year}{2019}\natexlab{}.
\newblock \showarticletitle{{Multi}-{Agent} {Pathfinding}: {Definitions},
  {Variants}, and {Benchmarks}}. In \bibinfo{booktitle}{\emph{International
  Symposium on Combinatorial Search}}, Vol.~\bibinfo{volume}{10}.
  \bibinfo{pages}{151--158}.
\newblock


\bibitem[\protect\citeauthoryear{Su, Adams, and Beling}{Su
  et~al\mbox{.}}{2021}]%
        {su2021value}
\bibfield{author}{\bibinfo{person}{Jianyu Su}, \bibinfo{person}{Stephen Adams},
  {and} \bibinfo{person}{Peter Beling}.} \bibinfo{year}{2021}\natexlab{}.
\newblock \showarticletitle{{Value}-{Decomposition} {Multi}-{Agent}
  {Actor}-{Critics}}. In \bibinfo{booktitle}{\emph{AAAI Conference on
  Artificial Intelligence}}, Vol.~\bibinfo{volume}{35}.
\newblock


\bibitem[\protect\citeauthoryear{Sukhbaatar, Lin, Kostrikov, Synnaeve, Szlam,
  and Fergus}{Sukhbaatar et~al\mbox{.}}{2018}]%
        {sukhbaatar2018intrinsic}
\bibfield{author}{\bibinfo{person}{Sainbayar Sukhbaatar},
  \bibinfo{person}{Zeming Lin}, \bibinfo{person}{Ilya Kostrikov},
  \bibinfo{person}{Gabriel Synnaeve}, \bibinfo{person}{Arthur Szlam}, {and}
  \bibinfo{person}{Rob Fergus}.} \bibinfo{year}{2018}\natexlab{}.
\newblock \showarticletitle{{Intrinsic} {Motivation} and {Automatic}
  {Curricula} via {Asymmetric} {Self}-{Play}}. In
  \bibinfo{booktitle}{\emph{ICLR}}.
\newblock


\bibitem[\protect\citeauthoryear{Sunehag, Lever, Gruslys, Czarnecki,
  et~al\mbox{.}}{Sunehag et~al\mbox{.}}{2018}]%
        {sunehag2017value}
\bibfield{author}{\bibinfo{person}{Peter Sunehag}, \bibinfo{person}{Guy Lever},
  \bibinfo{person}{Audrunas Gruslys}, \bibinfo{person}{Wojciech~Marian
  Czarnecki}, {et~al\mbox{.}}} \bibinfo{year}{2018}\natexlab{}.
\newblock \showarticletitle{{Value}-{Decomposition} {Networks} for
  {Cooperative} {Multi}-{Agent} {Learning} based on {Team} {Reward}}. In
  \bibinfo{booktitle}{\emph{AAMAS (Extended Abstract)}}.
\newblock


\bibitem[\protect\citeauthoryear{Sutton, McAllester, Singh, and Mansour}{Sutton
  et~al\mbox{.}}{2000}]%
        {sutton2000policy}
\bibfield{author}{\bibinfo{person}{Richard~S Sutton}, \bibinfo{person}{David~A
  McAllester}, \bibinfo{person}{Satinder~P Singh}, {and}
  \bibinfo{person}{Yishay Mansour}.} \bibinfo{year}{2000}\natexlab{}.
\newblock \showarticletitle{{Policy} {Gradient} {Methods} for {Reinforcement}
  {Learning} with {Function} {Approximation}}. In
  \bibinfo{booktitle}{\emph{Advances in Neural Information Processing
  Systems}}. \bibinfo{pages}{1057--1063}.
\newblock


\bibitem[\protect\citeauthoryear{Tan}{Tan}{1993}]%
        {tan1993multi}
\bibfield{author}{\bibinfo{person}{Ming Tan}.} \bibinfo{year}{1993}\natexlab{}.
\newblock \showarticletitle{{Multi}-{Agent} {Reinforcement} {Learning}:
  {Independent} vs. {Cooperative} {Agents}}. In \bibinfo{booktitle}{\emph{10th
  International Conference on Machine Learning}}. \bibinfo{pages}{330--337}.
\newblock


\bibitem[\protect\citeauthoryear{Tesauro et~al\mbox{.}}{Tesauro
  et~al\mbox{.}}{1995}]%
        {tesauro1995temporal}
\bibfield{author}{\bibinfo{person}{Gerald Tesauro} {et~al\mbox{.}}}
  \bibinfo{year}{1995}\natexlab{}.
\newblock \showarticletitle{{Temporal} {Difference} {Learning} and
  {TD}-{Gammon}}.
\newblock \bibinfo{journal}{\emph{Commun. ACM}} \bibinfo{volume}{38},
  \bibinfo{number}{3} (\bibinfo{year}{1995}), \bibinfo{pages}{58--68}.
\newblock


\bibitem[\protect\citeauthoryear{Vinyals, Babuschkin, Czarnecki, Mathieu,
  et~al\mbox{.}}{Vinyals et~al\mbox{.}}{2019}]%
        {vinyals2019grandmaster}
\bibfield{author}{\bibinfo{person}{Oriol Vinyals}, \bibinfo{person}{Igor
  Babuschkin}, \bibinfo{person}{Wojciech~M Czarnecki},
  \bibinfo{person}{Mathieu}, {et~al\mbox{.}}} \bibinfo{year}{2019}\natexlab{}.
\newblock \showarticletitle{{Grandmaster} {Level} in {StarCraft} {II} using
  {Multi}-{Agent} {Reinforcement} {Learning}}.
\newblock \bibinfo{journal}{\emph{Nature}} (\bibinfo{year}{2019}),
  \bibinfo{pages}{1--5}.
\newblock


\bibitem[\protect\citeauthoryear{Wang, Ren, Liu, Yu, and Zhang}{Wang
  et~al\mbox{.}}{2020}]%
        {wang2020qplex}
\bibfield{author}{\bibinfo{person}{Jianhao Wang}, \bibinfo{person}{Zhizhou
  Ren}, \bibinfo{person}{Terry Liu}, \bibinfo{person}{Yang Yu}, {and}
  \bibinfo{person}{Chongjie Zhang}.} \bibinfo{year}{2020}\natexlab{}.
\newblock \showarticletitle{{QPLEX}: {Duplex} {Dueling} {Multi}-{Agent}
  {Q}-{Learning}}. In \bibinfo{booktitle}{\emph{ICLR}}.
\newblock


\bibitem[\protect\citeauthoryear{Wang, Xiang, Huang, and Sartoretti}{Wang
  et~al\mbox{.}}{2023}]%
        {wang2023scrimp}
\bibfield{author}{\bibinfo{person}{Yutong Wang}, \bibinfo{person}{Bairan
  Xiang}, \bibinfo{person}{Shinan Huang}, {and} \bibinfo{person}{Guillaume
  Sartoretti}.} \bibinfo{year}{2023}\natexlab{}.
\newblock \showarticletitle{{SCRIMP}: {Scalable} {Communication} for
  {Reinforcement}-and {Imitation}-{Learning}-{Based} {Multi}-{Agent}
  {Pathfinding}}. In \bibinfo{booktitle}{\emph{2023 International Conference on
  Autonomous Agents and Multiagent Systems}}. \bibinfo{pages}{2598--2600}.
\newblock


\bibitem[\protect\citeauthoryear{Watkins and Dayan}{Watkins and Dayan}{1992}]%
        {watkins1992q}
\bibfield{author}{\bibinfo{person}{Christopher~JCH Watkins} {and}
  \bibinfo{person}{Peter Dayan}.} \bibinfo{year}{1992}\natexlab{}.
\newblock \showarticletitle{{Q}-{Learning}}.
\newblock \bibinfo{journal}{\emph{Machine Learning}} \bibinfo{volume}{8},
  \bibinfo{number}{3-4} (\bibinfo{year}{1992}), \bibinfo{pages}{279--292}.
\newblock


\bibitem[\protect\citeauthoryear{Wu, Yang, Hao, Hao, Zheng, Wang, and
  Taylor}{Wu et~al\mbox{.}}{2023}]%
        {wu2023portal}
\bibfield{author}{\bibinfo{person}{Jizhou Wu}, \bibinfo{person}{Tianpei Yang},
  \bibinfo{person}{Xiaotian Hao}, \bibinfo{person}{Jianye Hao},
  \bibinfo{person}{Yan Zheng}, \bibinfo{person}{Weixun Wang}, {and}
  \bibinfo{person}{Matthew~E Taylor}.} \bibinfo{year}{2023}\natexlab{}.
\newblock \showarticletitle{{PORTAL}: {Automatic} {Curricula} {Generation} for
  {Multiagent} {Reinforcement} {Learning}}. In \bibinfo{booktitle}{\emph{AAMAS
  (Extended Abstract)}}. \bibinfo{pages}{2460--2462}.
\newblock


\bibitem[\protect\citeauthoryear{Yu, Velu, Vinitsky, Gao, Wang, Bayen, and
  Wu}{Yu et~al\mbox{.}}{2022}]%
        {yu2022surprising}
\bibfield{author}{\bibinfo{person}{Chao Yu}, \bibinfo{person}{Akash Velu},
  \bibinfo{person}{Eugene Vinitsky}, \bibinfo{person}{Jiaxuan Gao},
  \bibinfo{person}{Yu Wang}, \bibinfo{person}{Alexandre Bayen}, {and}
  \bibinfo{person}{Yi Wu}.} \bibinfo{year}{2022}\natexlab{}.
\newblock \showarticletitle{{The} {Surprising} {Effectiveness} of {PPO} in
  {Cooperative} {Multi}-{Agent} {Games}}.
\newblock \bibinfo{journal}{\emph{Advances in Neural Information Processing
  Systems}}  \bibinfo{volume}{35} (\bibinfo{year}{2022}).
\newblock


\bibitem[\protect\citeauthoryear{Zhao, Zhuang, Huang, and Liu}{Zhao
  et~al\mbox{.}}{2023}]%
        {zhao2023curriculum}
\bibfield{author}{\bibinfo{person}{Cheng Zhao}, \bibinfo{person}{Liansheng
  Zhuang}, \bibinfo{person}{Yihong Huang}, {and} \bibinfo{person}{Haonan Liu}.}
  \bibinfo{year}{2023}\natexlab{}.
\newblock \showarticletitle{{Curriculum} {Learning} {Based} {Multi}-{Agent}
  {Path} {Finding} for {Complex} {Environments}}. In
  \bibinfo{booktitle}{\emph{2023 International Joint Conference on Neural
  Networks (IJCNN)}}. IEEE, \bibinfo{pages}{1--8}.
\newblock


\end{thebibliography}



\end{document}